\documentclass[%
 aip,
amsmath,amssymb,
preprint,%
]{revtex4-1}

\usepackage{graphicx}% Include figure files
\usepackage{dcolumn}% Align table columns on decimal point
\usepackage{bm}% bold math
\usepackage{hyperref}
\hypersetup{
    colorlinks = true,
    urlcolor   = blue,
    citecolor  = blue,
    linkcolor = blue,
}
\usepackage[utf8]{inputenc}
\usepackage{csquotes}
\usepackage[T1]{fontenc}
\usepackage{mathptmx}
\usepackage{etoolbox}
\usepackage{color}
\usepackage[mathlines]{lineno}
\usepackage[table, dvipsnames]{xcolor}
\usepackage{rotating}
\usepackage[export]{adjustbox}
\usepackage{array}
\makeatletter
\def\@email#1#2{%
 \endgroup
 \patchcmd{\titleblock@produce}
  {\frontmatter@RRAPformat}
  {\frontmatter@RRAPformat{\produce@RRAP{*#1\href{mailto:#2}{#2}}}\frontmatter@RRAPformat}
  {}{}
}%
\makeatother
\begin{document}
\preprint{AIP/123-QED
}

\title[]{Identifying feedback directions in the mechanisms driving self-sustained thermoacoustic instability in a single-element rocket combustor}
% Force line breaks with \\
\author{Somnath De}
\email{somnathde.mec@gmail.com}
\affiliation{Department of Aerospace Engineering, Indian Institute of Technology  Madras, 600036, India}% 
\affiliation{Center of Excellence for studying Critical Transition in Complex Systems,Indian Institute of Technology  Madras, 600036, India}

\author{Praveen Kasthuri}
\affiliation{Department of Aerospace Engineering, Indian Institute of Technology  Madras, 600036, India}%
\affiliation{Center of Excellence for studying Critical Transition in Complex Systems,Indian Institute of Technology  Madras, 600036, India}

\author{Matthew E. Harvazinski}
\affiliation{Air Force Research Laboratory, Edwards AFB, CA, 93524}%

\author{Rohan Gejji}
\affiliation{School of Aeronautics and Astronautics, Purdue University, West Lafayette, Indiana 47907, USA}%

\author{William Anderson}
\affiliation{School of Aeronautics and Astronautics, Purdue University, West Lafayette, Indiana 47907, USA}%

\author{R. I. Sujith}%
%\email{sujith@iitm.ac.in}
\affiliation{Department of Aerospace Engineering, Indian Institute of Technology  Madras, 600036, India}% 
\affiliation{Center of Excellence for studying Critical Transition in Complex Systems,Indian Institute of Technology  Madras, 600036, India}
\email{sujith@iitm.ac.in}
\date{\today}% It is always \today, today,

\begin{abstract}
%(word count: 253)
The occurrence of high frequency (> 1000 Hz) thermoacoustic instability (TAI) sustained by mutual feedback among the acoustic field, heat release rate oscillations, and hydrodynamic oscillations poses severe challenges to the operation and structural integrity of rocket engines. Hence, quantifying the differing levels of feedback between these variables can help uncover the underlying mechanisms behind such high frequency TAI, enabling redesign of combustors to mitigate TAI. However, so far, no concrete method exists to decipher the varying levels of mutual feedback during high-frequency TAI. In the present study, we holistically investigate the mutual influence based on the spatiotemporal 
%behavior of
directionality among acoustic pressure, heat release rate, hydrodynamic and thermal oscillations during TAI of a single-element rocket engine combustor. Using symbolic transfer entropy (STE), we identify the spatiotemporal direction of feedback interactions between those primary variables when acoustic waves significantly emerge  
during TAI. 
%From the spatiotemporal analysis using STE,
We unveil the influence of vorticity dynamics at the fuel collar (or the propellant splitter plate) as the primary stimulant over the heat release rate fluctuations to rapidly amplify the amplitude of the acoustic field. Further, depending on the quantification of the degree of the mutual information (i.e., the net direction of information), 
%between the variables
we identify the switches in dominating the thermoacoustic driving between the variables during TAI, each representing a distinct mechanism of a thermoacoustic state. Additionally, from this quantification, we analyze the relative dominance of the variables and rank-order the mutual feedback according to their impact on driving TAI. 
 %\textcolor{red}{Thus, this STE-based study helps accurately understand the feedback directions between the variables in an autonomous thermoacoustically excited rocket combustor.}

\end{abstract}

\keywords{Single-element rocket combustor, Thermoacoustic instability, Longitudinal mode oscillations, Symbolic transfer entropy, Mutual influence, Spatiotemporal directionality.}

\maketitle

\begin{table*} % Table 1
\caption{List of acronyms}\label{Tab:acronyms}
\centering
%m{'width'}
\setlength\arrayrulewidth{1.5pt}
%\begin{tabular}{|p{2.3cm}|p{2.3cm}|p{2.3cm}|p{2.3cm}|p{2.3cm}|p{2.3cm}|}
%\begin{tabular}{|p{4cm}|p{4cm}|}
\begin{center}
\begin{tabular}{ c|c} 
%\begin{tabular}{|p{4cm}|p{4cm}|}
%\begin{tabular}{ |l|l|l|l|l|l| }
\hline
%\begin{flushleft}Geometric configuration\end{flushleft}&\begin{flushleft} measurements \end{flushleft}\\
%\textbf{Opening conditions}&\textbf{Values}\\
%\hline
\textbf{CVRC} & Continuously variable resonance combustor\\
%\hline
\textbf{FFT}&  Fast Fourier transform\\
%\hline
\textbf{HALCO} &  High amplitude limit cycle oscillations\\
%\hline
\textbf{HAMSTER} &  Hydrodynamic and mixing study of a thermoacoustically excited rocket\\
%\hline
\textbf{LES} &  Large Eddy Simulation\\
%\hline
%\textbf{m} &  Meter\\
%\hline
%\textbf{mm} &  Millimeter\\
%\hline
%\textbf{ms} &  Millisecond\\
%\hline
\textbf{SDF} &  Symbolic dynamic filtering\\
%\hline
\textbf{STE} & Symbolic transfer entropy\\
%\hline
\textbf{TAI}&   Thermoacoustic instability \\
\hline
%Combustion chamber diameter&  2''\\
%\hline
\end{tabular}
\end{center}
%\
\end{table*}

\begin{table*} % Table 2
\caption{Nomenclature}\label{Tab:nomenclature}
\centering
%m{'width'}
\setlength\arrayrulewidth{1.5pt}
%\begin{tabular}{|p{2.3cm}|p{2.3cm}|p{2.3cm}|p{2.3cm}|p{2.3cm}|p{2.3cm}|}
%\begin{tabular}{|p{4cm}|p{4cm}|}
%\begin{center}
%\begin{tabular}{ c|c} 
\setlength\extrarowheight{-4pt}
\begin{tabular}{|p{6.22cm}|p{6.22cm}|}
%\begin{tabular}{ |l|l|l|l|l|l| }
\hline
%\begin{flushleft}Geometric configuration\end{flushleft}&\begin{flushleft} measurements \end{flushleft}\\
%\textbf{Opening conditions}&\textbf{Values}\\
%\hline
\textbf{Physical variables} & \textbf{Parameter}\\
%\hline
${p}^{\prime}$: acoustic pressure oscillations (in kPa) &  $\Phi$ : Equivalence ratio\\
%\hline
${\omega}$ : Vorticity (in 1/s) &  \textbf{Measures}\\
%\hline
$\dot{Q}^{\prime}$: Heat release rate oscillations (in W/m$^{3}$) & $T^{S}_{{p^{\prime}}{\rightarrow}{\theta^{\prime}}}$: Symbolic transfer entropy from $p^{\prime}$ to $\theta^{\prime}$  \\
%\hline
${u}^{\prime}$: flow velocity oscillations (in m/s) &  $\Delta S_{{p^{\prime}}{\rightarrow}{I^{\prime}_{OH*}}}$: Directionality index calculated in the direction of ${p^{\prime}}{\rightarrow}{I^{\prime}_{OH*}}$\\
%\hline
%\textbf{m} &  Meter\\
%\hline
%\textbf{mm} &  Millimeter\\
%\hline
%\textbf{ms} &  Millisecond\\
%\hline
${\theta}^{\prime}$: combustion temperature oscillations (in Kelvin) & $\Delta S_{{p^{\prime}}{\rightarrow}{\dot{Q}^{\prime}}}$: Directionality index calculated in the direction of ${p^{\prime}}{\rightarrow}{\dot{Q}^{\prime}}$  \\
%\hline
$I^{\prime}_{OH*}$: OH* chemiluminescence oscillations (a.u.) & $\Delta S_{{p^{\prime}}{\rightarrow}{\omega}}$: Directionality index calculated in the direction of ${p^{\prime}}{\rightarrow}{\omega}$ \\
%\hline
\textbf{Measures}&  \\
$T^{S}_{{p^{\prime}}{\rightarrow}{I^{\prime}_{OH*}}}$: Symbolic transfer entropy from $p^{\prime}$ to $I^{\prime}_{OH*}$ & $\Delta S_{{p^{\prime}}{\rightarrow}{u^{\prime}}}$: Directionality index calculated in the direction of ${p^{\prime}}{\rightarrow}{u^{\prime}}$ \\
$T^{S}_{{p^{\prime}}{\rightarrow}{\dot{Q}^{\prime}}}$: Symbolic transfer entropy from $p^{\prime}$ to $\dot{Q}^{\prime}$ & $\Delta S_{{p^{\prime}}{\rightarrow}{\theta^{\prime}}}$: Directionality index calculated in the direction of ${p^{\prime}}{\rightarrow}{\theta^{\prime}}$  \\
$T^{S}_{{p^{\prime}}{\rightarrow}{\omega}}$: Symbolic transfer entropy from $p^{\prime}$ to $\omega$ & $\Delta S_{{\dot{Q}^{\prime}}{\rightarrow}{\omega}}$: Directionality index calculated in the direction of ${\dot{Q}^{\prime}}{\rightarrow}{\omega}$  \\
$T^{S}_{{\omega}{\rightarrow}{\dot{Q}^{\prime}}}$: Symbolic transfer entropy from $\omega$ to $\dot{Q}^{\prime}$ & $\langle\Delta S_{{p^{\prime}}{\rightarrow}{\dot{Q}^{\prime}}}\rangle$: Area averaged directionality index between $p^{\prime}$ and $\dot{Q}^{\prime}$  \\
$T^{S}_{{p^{\prime}}{\rightarrow}{u^{\prime}}}$: Symbolic transfer entropy from $p^{\prime}$ to $u^{\prime}$ & $\langle\Delta S_{{p^{\prime}}{\rightarrow}{\omega}}\rangle$: Area averaged directionality index between $p^{\prime}$ and $\omega$  \\
$\langle\Delta S_{{p^{\prime}}{\rightarrow}{u^{\prime}}}\rangle$: Area averaged directionality index between $p^{\prime}$ and $u^{\prime}$ & $\langle\Delta S_{{p^{\prime}}{\rightarrow}{\theta^{\prime}}}\rangle$: Area averaged directionality index between $p^{\prime}$ and $\theta^{\prime}$ \\
$\langle\Delta S_{{\dot{Q}^{\prime}}{\rightarrow}{\omega}}\rangle$: Area averaged directionality index between $\dot{Q}^{\prime}$ and $\omega$ & \\
\hline
%Combustion chamber diameter&  2''\\
%\hline
\end{tabular}
%\end{center}
%\
\end{table*}

\section{Introduction}\label{sec:introduction}

Rockets are indispensable for placing satellites in the desired orbit for various essential applications ranging from weather forecasting, GPS tracking, telecommunications, environmental monitoring, and military applications to tracking the movement of space objects~\cite{taylor2017introduction}. They also enable the innate human desire for space exploration. Hence, there is no doubt that rockets are one of the greatest innovations of scientists.

The occurrence of thermoacoustic instability (TAI)~\cite{juniper2018sensitivity,pawar2017thermoacoustic} offers considerable hurdles to engineers in design and safe operation of rocket engine combustors across the mission envelope. TAI arises in the rocket engine combustor when a two-way feedback between the acoustic field developed in the combustor, heat release rate caused by combustion and hydrodynamic field~\cite{sujith2021thermoacoustic,lieuwen2021unsteady}is established at different temporal and spatial scales. Under certain conditions which are not fully understood, these mutual interactions lead to ruinously large amplitude oscillations of acoustic pressure in rocket engine combustor~\cite{crocco1956theory,culick1995overview,yang1995liquid}. The self-sustaining nature of TAI can create excessive vibration, resulting in structural damage to 
%components
the engine and catastrophic failure of electrical components in the launch vehicle and its payload ~\cite{culick2006unsteady}. TAI can also induce excessive heat transfer and overwhelm the thermal protection system, resulting in thermal distress and eventual failure~\cite{schob1963experimental}. In gas turbine combustors, TAI often precedes flame blowout~\cite{de2020application,de2021early,unni2016precursors} and flashback ~\cite{kabiraj2012nonlinear}. In the case of a rocket, the situation can often be serious as mission failures can occur~\cite{sujith2021thermoacoustic}. A well-known instance is the multiple challenges faced in the development of the well-known F-1 engine used for Apollo space missions~\cite{yang1995liquid}. The program was delayed to understand and fix the repetitive failures due to TAI during testing and caused considerable financial loss.

Despite decades of research, TAI still poses significant challenges to the development of flight worthy rocket engines. Consequently, understanding the mechanism of TAI in rocket engines remains one of the focal research interests to the community~\cite{sirignano2015driving,urbano2017driving,nguyen2018longitudinal,armbruster2019injector,shima2021formation,kasthuri2022coupled,kasthuri2022investigation}. With this background, an in-depth understanding of the varying level of feedback interaction between the primary variables can be of the utmost importance to designing and potentially developing predictive and mitigation strategies for TAI in rocket engine combustors.       
  
Conventionally, rocket engines are designed with several flow injectors connected to a common dump volume, where vortex shedding occurs behind a backward-facing step at the injector face~\cite{harvazinski2013analysis, huang2016analysis}. If the velocity of the flow upstream of the vortex shedding point changes, the vortex shedding frequency also can change. Further, the difference in temporal and spatial scales between the vortex shedding from fuel and oxidizer can regulate the 
%combustion mode
heat release rate oscillations downstream. The flow mechanism behind the backward-facing step (i.e., a flame holder) can modify the shear layer developed from the injector. On the other hand, the acoustic wave established in the combustor can modify the flow dynamics both upstream and downstream of the backward-facing step. Earlier, Culick and co-workers~\cite{culick1994some,jahnke1994application} theoretically explained that acoustic oscillations grow in a combustion chamber until limited by the nonlinear process. Different factors, such as the injection and combustion of propellants at near-critical and supercritical thermodynamic conditions, turbulent flow, flow separation at sharp edges, rapid flow expansions, shock discontinuities, extreme rate of heat addition, etc.,  yield these nonlinearities~\cite{sirignano2015driving}. Hence, flame dynamics are greatly affected by the nonlinear interaction process among physical variables, such as acoustic, combustion, and hydrodynamic processes, and the thermodynamic properties of the reacting flow~\cite{culick1970stability,lieuwen2021unsteady}.
%Therefore, only a nonlinear theory can predict the complex wave profile of acoustic pressure carrying shock discontinuities, while a continuous sinusoidal profile can be tracked by linear theory~\cite{sirignano2015driving}}.  
Due to the complex interaction between the variables and the presence of nonlinearities, a linear theory is often inadequate to explain the temporal and spatial dynamics of TAI occurring inside rocket combustors~\cite{o2015transverse,acharya2018non}. Consequently, theories that account for the nonlinearities of the reacting flow inside the rocket engine combustor and the mechanisms of TAI have since been devised. 

Several studies explored the behavior of TAI using different nonlinear methods. In the earlier stage, Priem \& Guentert~\cite{priem1962combustion} attempted to investigate the behavior and stability of finite amplitude waves in a liquid-propelled rocket engine (LPRE) combustors assuming the wave oscillations to be one dimensional. In a subsequent study, Sirignano~\cite{sirignano1964theoretical} extended that study using various theoretical, numerical and analytical solution methods considering the wave oscillations in two dimensions. After that, scientists extensively investigated the nonlinear stability characteristics of LPRE combustors by using the concept of time lag ($\tau$) defined by the period required for processes to occur before the final combustion~\cite{crocco1956theory}, an index ($n$) measuring the interaction between different processes in the combustion zone~\cite{zinn1968theoretical}, and a modified Galerkin method~\cite{zinn1971application}. Using the Galerkin approach~\cite{zinn1971application}, it was discovered that a state of TAI exhibits shock-like behavior, as determined by the characterization of the unsteady combustion process. The predicted TAI behavior using the nonlinear equations agreed well with the experimental results. However, these analytical models are limited to studying the unsteady combustion of moderate amplitude and low Mach number flows. Also, the Galerkin methods necessitate prior knowledge of the relevant oscillation modes~\cite{sirignano2013two, popov2014stochastic}. These reduced-basis techniques have been shown to be inadequate if the chosen fundamental functions do not include natural modes, their harmonics, and specific sub-harmonics~\cite{sirignano2013two}.

With the progress of high-efficiency computing, numerical simulation has been discovered to be a valuable addition to the analytical solution and experimental data. Scientists provided notable efforts in understanding TAI in liquid rocket engine combustors by modeling the computational flow domain using large eddy simulation (LES)~\cite{selle2014prediction,urbano2016exploration,urbano2017driving,nguyen2018longitudinal}. Selle et al.~\cite{selle2014prediction} performed a rigorous analysis on the prediction of the occurrence of TAI using an LES solver for better mimicking the flame dynamics and a Helmholtz solver for solving the acoustic field. They recommended LES as a promising approach for future developments in numerical simulations of rocket engine combustors. Using the LES dataset, Urbano et al.~\cite{urbano2017driving} found that flames from the injectors drive the acoustic field, while the coupling between acoustic and hydrodynamic field acts as the damping mechanism of TAI in the transverse mode dominated rocket engine combustor. Nguyen et al.~\cite{nguyen2018longitudinal} conducted an investigation on TAI in LPRE by modeling the turbulence using a hybrid Reynolds-averaged Navier–Stokes/LES approach, while the interaction between combustion and turbulence was modeled using the flamelet/progress-variable approach. Further, a group from the German Aerospace Center (Deutsches Zentrum für Luft- und Raumfahrt, abbreviated as DLR) attempted to understand the coupling between acoustic and combustion fields in tangential mode-dominated rocket engine combustors using modal decomposition methods~\cite{hardi2016approaches}.
 %These numerical studies shed light on TAI investigation to some extent.

Additionally, several experimental and numerical studies on gas turbine~\cite{kabiraj2012nonlinear,nair2014multifractality,murugesan2015combustion,godavarthi2018coupled,murayama2019attenuation} and liquid rocket engine combustors~\cite{hashimoto2019spatiotemporal,kasthuri2019dynamical,kobayashi2019spatiotemporal,shima2021formation,kawano2023complex} provided significant efforts to understand the complex behavior of TAI under the framework of dynamical systems theory. Dynamical systems theory helps to gain new insights into the development of flame dynamics due to the interaction between the physical variables. Kabiraj and Sujith~\cite{kabiraj2012nonlinear} found that a subcritical Hopf bifurcation resulting from a nonlinear interaction between the acoustic field and a confined laminar flame can lead to TAI. In that study, subcritical Hopf bifurcation was observed when a sudden jump resulting in a single point in the bifurcation plot occurred at a particular flame location. 

Further, the recurrence approach from dynamics theory is found to be promising in demarcating different dynamical states prior to the onset of TAI~\cite{kabiraj2012nonlinear,unni2016precursors,godavarthi2018coupled,hernandez2019detection,kasthuri2020recurrence}. Recently, Unni and Sujith~\cite{unni2016precursors} revealed using recurrence plots that turbulent combustors undergo a state of intermittency before the onset of TAI. They also noticed that the Shannon entropy that measures the complexity becomes low during TAI since a regular structure in the temporal behavior of the combustor emerges during that time. Kasthuri et al.~\cite{kasthuri2019dynamical} found that a ratio between two recurrence measures, determinism, and recurrence rate, becomes high during aperiodic oscillations in a  liquid rocket engine combustor, while it decays to almost zero for the periodic oscillations (TAI).    
Further, recurrence theory combined with the synchronization theory~\cite{godavarthi2018coupled} can be useful in understanding the coupling behavior between acoustic pressure and heat release rate oscillations for a combustion system approaching TAI at which the physical variables show a generalized synchronization (GS). The investigations found that joint transitivity~\cite{godavarthi2018coupled}, transitivity ratio~\cite{godavarthi2018coupled}, and synchronization index~\cite{murayama2019attenuation} based on joint and inter-system recurrence can be promising measures to detect the onset of GS. Nevertheless, the characterization of the coupling between acoustic field and heat release rate fluctuation was the main emphasis of those earlier studies~\cite{godavarthi2018coupled,murayama2019attenuation}, with the direction of the feedback interaction between the variables receiving only relatively minor attention.      

Besides recurrence theory, the analysis of the transitions to an unstable state using complex network approaches gained considerable interest in the study of TAI~\cite{sujith2020complex}. The interactions between the components constitute a complex network in which the components are represented as nodes, and the pairwise interaction is considered as links. 
%Under this concept, 
In this framework of complex networks, Murugesan and Sujith~\cite{murugesan2015combustion} found that the low amplitude, aperiodic pressure fluctuations observed during the occurrence of combustion noise exhibit scale-free behavior. On the other hand, the scale-free behavior disappears and order emerges, leading to the formation of a regular network, during the transition to TAI~\cite{murugesan2015combustion}. Recent studies~\cite{krishnan2019emergence,shima2021formation} tried to understand the type of formation of thermoacoustic power production by constructing a network based on the product of acoustic and heat release rate fluctuations. Krishnan et al.~\cite{krishnan2019emergence} reported that the generation of acoustic power happens in an incoherent manner during a stable state, while that formation is seen to occur coherently during a formation of large-scale vorticity prior to TAI. Further, the study was extended to 
%recognize the dissimilarities in the 
understand the vorticity interaction in turbulent flow field during TAI using time-varying weighted spatial networks ~\cite{hashimoto2019spatiotemporal,krishnan2021suppression,mori2023feedback} in both gas turbine and liquid rocket engine combustors. Recently, Kasthuri et al.~\cite{kasthuri2022investigation} found a significant difference in the coherence structure of the flame fluctuations between an intermittent state and TAI using positive and negatively correlated weighted networks. They found that the highly correlated flame regions tend to connect with other areas, leading to a highly coherent flame oscillation during TAI.      

Although several studies~\cite{matsuyama2016large,hashimoto2019spatiotemporal,kasthuri2019dynamical,kasthuri2020recurrence} focused on the transition to TAI, the understanding of the driving mechanism, especially during TAI is 
%not clear 
still one of the main concerns in liquid rocket engine combustors. Additionally, the control of TAI becomes more difficult in the liquid engine rocket combustors due to unmixed propellants operating at near-stoichiometry with no diluents and high chamber pressure. A recent study~\cite{hashimoto2019spatiotemporal} 
%in \textcolor{blue}{liquid} rocket engine combustors 
employed the concept of transfer entropy to understand the direction of influence between the heat release rate and the acoustic field during a transition to a fully developed TAI in a liquid rocket engine combustor dominated by the first tangential mode. During TAI, many subsystems participate in the coupling process~\cite{lieuwen2021unsteady} and the directional nature of the coupling is not fully understood. Therefore, a holistic study is necessary to analyze the directional coupling amongst the physical variables. 

%However, since the mechanism of TAI in \textcolor{blue}{liquid} rocket engine combustors is not straightforward, a holistic study is necessary to analyze the interaction amongst the physical variables~\cite{lieuwen2021unsteady}. 

 Over the last two decades, research shows that symbolic dynamic filtering, which is a branch of nonlinear dynamics, has gained substantial recognition in various fields, including engineering~\cite{gupta2007symbolic,unni2015online}, astrophysics~\cite{schwarz1993analysis}, finance~\cite{xu2017symbolic}, medicine~\cite{kurths1995quantitative,beim2000symbolic}, climate science~\cite{du2011application} etc. However, a recent trend in this line of research shows that the concept of symbolic analysis combined with a variety of dynamical approaches becomes robust for understanding a variety of complex systems~\cite{shima2021formation,asami2022dynamic,mori2023feedback}. Within a few years, transfer entropy combined with the hypothesis of symbolic dynamics filtering has attracted significant attention in several disciplines, such as physics~\cite{staniek2008symbolic,martini2011inferring}, engineering~\cite{xie2019adaptive,bauer2004specifying}, medical science~\cite{staniek2008symbolic,chang2013symbolic, gao2018multichannel}, and transmission of diseases~\cite{kissler2020symbolic}. Compared to the well-established techniques such as Granger causality~\cite{granger1969investigating} and permutation entropy~\cite{bandt2002permutation}, symbolic transfer entropy (STE) can be used to reveal asymmetric information flows in stochastic environments with minimal observational noise. Recently, Shima et al.~\cite{shima2021formation} used this concept to examine the effect of upstream flow velocity fluctuations on the heat release rate oscillations and synchronized behavior between the acoustic and the heat release rate oscillations during a transition to TAI in a liquid rocket engine combustor with an off-center installed coaxial injector. However, this study primarily focused on the transition to TAI, with the aim of developing a prediction technique for determining the onset of TAI. 
 %Consequently, we do not have a clear understanding of the direction of feedback interactions between the sub-systems (physical variables) during the state of TAI for which the phenomenon always pose a challenge to the rocket engine developers.}

In light of the advancement of STE, we use this concept to understand the mutual influence between the fluctuations in heat release rate, acoustic, flow velocity, vorticity, and thermal behavior for 
%$CH_4-O_2$ combustion of a thermoacoustically excited 
a single-element liquid rocket engine combustor~\cite{harvazinski2020large, fuller2019dynamic}. Unlike the earlier works performed on liquid rocket engine combustors dominated by transverse and tangential modes~\cite{yang1995liquid, crocco1967nonlinear, hashimoto2019spatiotemporal,urbano2017driving, shima2021formation, mori2023feedback}, we work on simulation data corresponding to an operating condition when the first longitudinal mode (1L) is found to be dominant during TAI in the liquid rocket engine combustor. The high-fidelity spatiotemporal simulation data was generated at Purdue University using General Equation and Mesh Solver (GEMS) that can solve the coupled Navier-Stokes equations along with a single energy equation and the coupled species equations~\cite{lian2009solution}. The CFD dataset was further validated with experimental results in an earlier work~\cite{harvazinski2020large}. Such a comparison is helpful
for gaining confidence in the simulation results.

In the present study, the usage of CFD data, specifically LES, over experimental data allows us to examine a wide range of flow and reaction variables as opposed to just point measurements of acoustic pressure and, in some cases, chemiluminescence intensity of OH$^*$/CH$^*$ radicals. Analyzing this dataset, we quantify the level of mutual influence between the physical variables and identify the direction of the dominant influence between them on spatiotemporal scales. Based on an extensive understanding of the mechanism undergoing the mutual influences among the variables, we further calculate the degree of mutual influence (i.e., the net mutual influence between the variables), which helps to identify the relatively stronger influencing (or primary driving) variable and measure the influence of the stronger variable on the other variables. We find the vorticity to be  
the stronger influencing variable during amplification (reduction) of the acoustic pressure fluctuations to a local maxima (minima) and find seven distinct stages of the feedback interactions based on the variation in the ranking order of the variables during one cycle of TAI. In this study, we unveil that the variables with the dominant influence change, indicating a shift in the functioning of the liquid rocket engine combustors during TAI.

The rest of the paper is arranged as follows. We provide the details of a liquid rocket engine combustor that is known as HAMSTER combustor and discuss the operating conditions adopted to obtain TAI in Sec.~\ref{sec:combustor}. Then, we describe the method of STE used to understand the mutual influence that elucidates the feedback interaction between the variables in detail in Sec.~\ref{sec:methods}. Next, we analyze this method on the temporal and spatiotemporal data to understand the mechanism of TAI in a 1L-dominated HAMSTER combustor in Sec.~\ref{sec:results}. Finally, we summarize the significant findings in Sec.~\ref{sec:conclusions}.

%The rest of the paper is arranged as follows. We provide \textcolor{blue}{the details} of the rocket engine combustor and \textcolor{blue}{discuss the operating conditions adopted to obtain TAI} in Sec.~\ref{sec:combustor}. Then, we describe the method of STE \textcolor{blue}{used to understand the mutual influence that enlightens the feedback interaction between the variables} in detail in Sec.~\ref{sec:methods}. Next, we analyze this method on the temporal and spatiotemporal data to understand the mechanism of TAI in a 1L-dominated rocket engine combustor in Sec.~\ref{sec:results}. Finally, we summarize the significant findings in Sec.~\ref{sec:conclusions}.    

\section{Details of the liquid rocket engine combustor:}\label{sec:combustor}

\begin{figure*}[ht!]%
\centering
\includegraphics[width=0.8\textwidth]{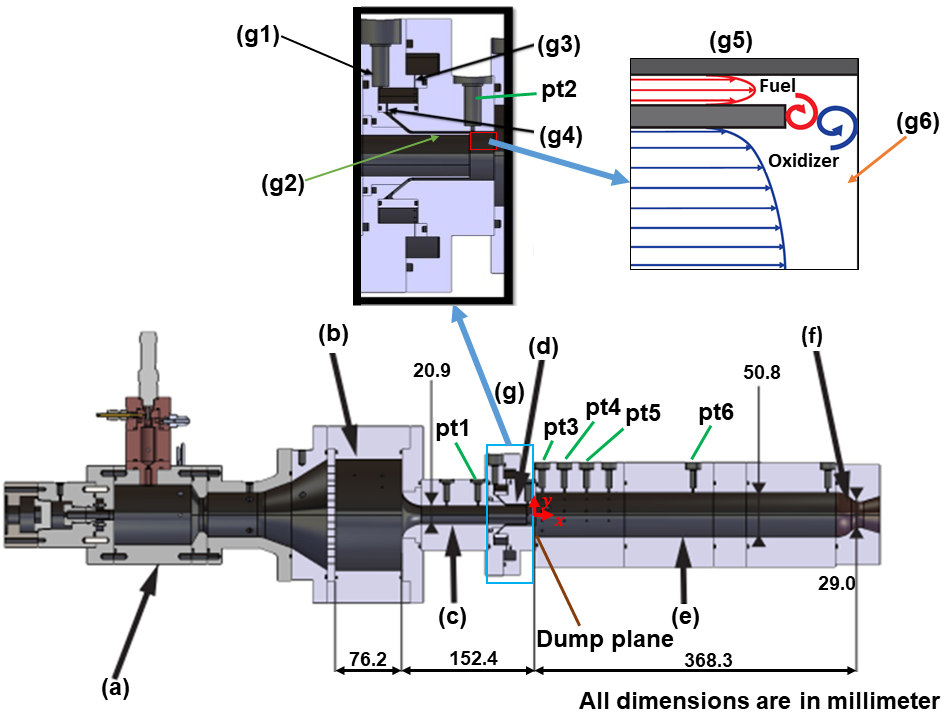}
\caption{A schematic diagram of the HAMSTER combustor~\cite{harvazinski2020large,fuller2019dynamic} is illustrated. This single-element combustor consists of preburner chamber (a), oxidizer manifold (b), oxidizer post (c), fuel injector (d), combustion chamber (e) and nozzle (f). The labeled subsections of the zoomed view of g are: fuel manifold (g1), fuel collar or propellant splitter plate (g2), inner/outer manifold choke point (g3), fuel manifold choke (g4), converging portion of fuel annulus (g5), and injector recess (g6). The locations of the pressure transducers used in this study are indicated by pt1-pt6.}\label{fig_setup}
\end{figure*}

%\textcolor{red}{(This section would most likely be rewritten completely by Rohan and Matthew when it goes to them.)}

We represent the schematic of the combustor used in an investigation of HAMSTER (hydrodynamic and mixing study of thermoacoustically excited rocket) experiments conducted at Purdue University~\cite{fuller2019dynamic} in Fig.~\ref{fig_setup}. The design and conduction of the investigations of HAMSTER are influenced by the study of Continuously-Variable Resonance Combustor (CVRC)~\cite{yu2012spontaneous}. The limitations and learning acquired from the CVRC simulation and experiments served as a foundation for the new design's advancement and improvement in HAMSTER. The configuration of HAMSTER uses a single element shear coaxial injector similar to those found in modern high-performance rocket engines. As shown in Fig.~\ref{fig_setup}, the methane-oxygen rocket combustor rig primarily comprises a pre-burner tube (a), oxidizer manifold (b), oxidizer post (c), fuel injector (d), combustion chamber (e) and nozzle (f). The $H_2/O_2$  preburner is used to warm the oxidizer (composition: 94.8$\%$ O$_{2}$ and 5.2$\%$  H$_{2}$). 
%Thus, oxidizer rich preburner injects warm O2 to the oxidizer manifold. 
In the present study, the oxidizer is preheated to 705 K.

%HAMSTER is designed in such a way that the acoustic responses throughout the system can be characterized for each subsection of the combustor. 
The oxidizer post inlet is designed to produce a smooth flow towards the fuel injector. It features a moderate curvature to prevent the injection of oxidizer from being hydrodynamically perturbed. A schematic of the fuel injector (g) is shown in Fig.~\ref{fig_setup} through different subsections, namely: fuel manifold (g1), fuel collar or fuel-oxidizer splitter plate (g2), inner/outer manifold choke point (g3), and fuel manifold choke (g4). The fuel flow is accelerated for the area reduction caused by the converging portion of the fuel annulus (see g5), which also aids in settling the flow.  
%The flexibility of the fuel injector allows for either inclusion or exclusion of fuel manifold participation in the combustion. This also makes a scope to conduct research to determine the impact of fuel flow manipulation on combustion stability.
The length of the annulus is adequate for the fuel flow to develop before it enters the injector recess (g6). The flow of the propellants before entering into the recess is shown in the g5 section of Fig.~\ref{fig_setup}.   
The propellant flow distribution plays a crucial role in each manifold. 

In this analysis, gaseous propellants, methane ($CH_4$) and oxygen ($O_2$), are used to avoid the requirement of model atomization. Further, one can accurately model the chemical kinetics of $CH_4$ and $O_2$ in the simulation of the combustor. Therefore, a previous study~\cite{harvazinski2020large} showed a good agreement between experimental and simulated results using these propellants. In the simulation of the combustor, GRI Mech 1.2~\cite{mechgri} is used to model chemical kinetics. The current study treats the propellants and combustion products as ideal gasses.

The simulations are conducted using GEMS, an in-house Purdue code. The code involves an intricately coupled Navier-Stokes solver with second order precision in time and space. The method for modelling turbulence is detached eddy simulations, which collect large-scale motions up to the grid length scale, while the sub-grid turbulence is modeled with a $k-\omega$ model~\cite{wilcox2008formulation}. NASA polynomial coefficients~\cite{mcbride1993coefficients} are used to derive the thermal and transport properties. For the discretization, the source terms, fluxes, and boundary conditions are handled implicitly. The dual time stepping approach is used to minimise the estimated factorization errors and acquire the solution of the linearized equations by the use of the line Gauss-Seidel algorithm.

%Further, 
Table~\ref{Tab:1} provides a detailed description of the operating conditions of the propellants chosen for simulation. In the current study, we use higher-fidelity simulation data obtained by using a high spatially and temporally resolved three-dimensional flow domain at a chamber pressure of 12 bars during the state of thermoacoustic instability (TAI). This LES dataset has been benchmarked with the corresponding experimental data by comparing the average and RMS pressure at different sections of the rocket combustor and dominant frequencies at the head end (i.e., dump region) of the combustor~\cite{harvazinski2020large}. A previous study of the Purdue group~\cite{harvazinski2013analysis} found that the amplitude of the first longitudinal mode can be captured by two-dimensional axisymmetric simulations utilizing both coarse and fine grid to within an order of magnitude of the experimental measurement. Following this, we consider a two-dimensional (\textit{$x$}, \textit{$y$}) flow domain by considering \textit{$z$} = 0 in our analysis. In Sec.~IV,
%In Sec.~\ref{sec:results}, 
we shift the \textit{$x-y$} coordinate at the dump plane. Hence, in our analysis, \textit{$x$} shows positive values towards the downstream or right side of the combustion chamber, while that becomes negative towards the upstream or left side of the combustion chamber. \textit{$y$} shows the positive (negative) values towards the upward (downward) direction from the centerline of the combustor. Further details of the LES model, combustor hardware, and operating conditions are mentioned in Harvazinski et al. \cite{harvazinski2020large}.  

%The results of calculations demonstrate that the amplitude of the first longitudinal mode may be captured by two-dimensional axisymmetric simulations utilising both coarse and fine grids to within an order of magnitude of the experiment's measurements.

%In the next section, we discuss the computational approach taken for simulation study of HAMSTER~\cite{harvazinski2020large}.

\begin{table*} % Table 3
\caption{Operating conditions taken for investigating HAMSTER in experiment and simulation at Purdue University~\cite{fuller2019dynamic,harvazinski2020large} are shown in the table.}\label{Tab:1}
\centering
%m{'width'}
\setlength\arrayrulewidth{1.5pt}
%\begin{tabular}{|p{2.3cm}|p{2.3cm}|p{2.3cm}|p{2.3cm}|p{2.3cm}|p{2.3cm}|}
%\begin{tabular}{|p{4cm}|p{4cm}|}
\begin{center}
\begin{tabular}{ c|c} 
%\begin{tabular}{|p{4cm}|p{4cm}|}
%\begin{tabular}{ |l|l|l|l|l|l| }
\hline
%\begin{flushleft}Geometric configuration\end{flushleft}&\begin{flushleft} measurements \end{flushleft}\\
\textbf{Opening conditions}&\textbf{Values}\\
\hline
Oxidizer temperature & 705 K\\
%\hline
Equivalence ratio ($\Phi$)&   0.8\\
%\hline
Fuel temperature &  296 K\\
%\hline
Oxidizer flow rate ($\dot{Q}_{Ox})$& 0.425 kg/s\\
%\hline
Fuel  flow rate ($\dot{Q}_{F}$)&  0.0884 kg/s\\
\hline
%Combustion chamber diameter&  2''\\
%\hline
\end{tabular}
\end{center}
%\
\end{table*}

%In order to extract and rank-order the various feedback directions among the primary variables governing TAI, we use the concept of symbolic transfer entropy~\cite{staniek2008symbolic} that relies on the combination of the information theory and symbolic dynamic filtering. Symbolic transfer entropy is useful for examining the directional information flow between two physical variables. To calculate the transfer entropy between two physical variables, we use both joint probability and conditional probability between the rank order patterns of those variables. Further, introducing a time delay in calculating those probabilities helps better to understand the correlation between the variables~\cite{staniek2008symbolic}. In other words, conditional probability in transfer entropy calculations aids us in extracting the asymmetric flow information between the variables. Thus, by calculating the transfer entropies between two variables, we can extract the information on the flow of influences that one variable has on the other. Furthermore, a net flow of the entropy, which is also regarded as the directionality or degree of mutual information,  can directly identify the dominant feedback pathways during \textcolor{blue}{TAI}. The proposed approach enables identification of the driving variables that play a crucial role in sustaining \textcolor{blue}{the limit cycle oscillations during the state of TAI in rocket engine combustors.

\section{Methodology:}\label{sec:methods}

In order to extract and rank-order the various feedback directions among the primary variables governing TAI, we use the concept of symbolic transfer entropy~\cite{staniek2008symbolic} (STE) that integrates information theory~\cite{kaiser2002information} and symbolic dynamic filtering (SDF)~\cite{rao2009review}. Symbolic transfer entropy is helpful for examining the directional information flow between two physical variables.

\begin{figure*}[ht!]%
\centering
\includegraphics[width=0.98\textwidth]{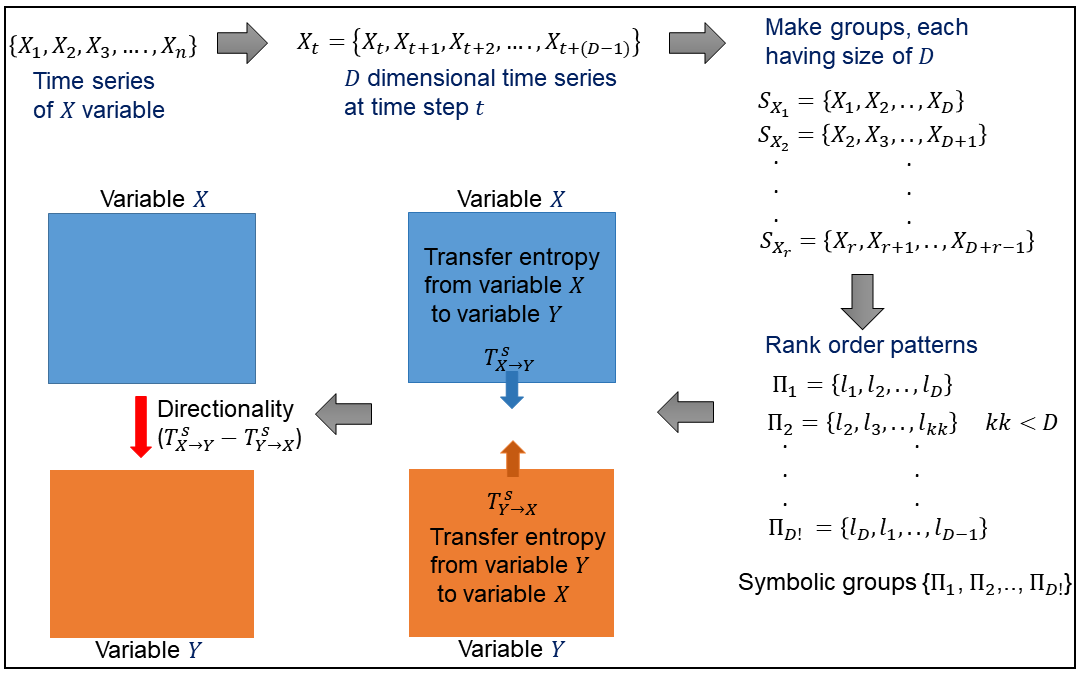}
\caption{The concept of symbolic transfer entropy is illustrated here. Based on the rank order patterns of the time series, we obtain transfer entropy of one variable ($X$) to another variable ($Y$).}\label{fig_STEmethodology}
\end{figure*}

\subsection{Concept of transfer entropy}
The concept of transfer entropy~\cite{staniek2008symbolic} (TE) from information theory pertains to the information flow between two subsystems of a system. For instance, we compute the following expression to determine how much $X$ has an impact on $Y$ ($T_{X{\rightarrow}Y}$)~\cite{bandt2002permutation}:
\begin{equation}
    T_{X{\rightarrow}Y} = \sum_{n} P(Y(t_{n+1}),Y(t_n),X(t_n))log_{2}\frac{P(Y(t_{n+1})|Y(t_n),X(t_n))}{P(Y(t_{n+1})|Y(t_n))}
\label{eq:TE}
\end{equation}
Here, $Y(t_{n+1})$ and $X(t_n)$ are the values of $Y$ and $X$ at time steps $t_{n+1}$ and $t_n$ respectively. Besides, $P(Y(t_{n+1}),Y(t_n),X(t_n))$ is the joint probability of $Y(t_{n+1})$, $Y(t_n)$ and $X(t_n)$. $P(Y(t_{n+1})|Y(t_n))$ is the conditional probability of $Y(t_{n+1})$ given $Y(t_n)$.   
The traditional transfer entropy calculation deals with a issue related to the type of dataset as the inputs from the various operating fields are not identical ~\cite{hashimoto2019spatiotemporal}. However, combining with symbolic dynamic filtering addresses that problem by performing the ranking order patterns of the data from various fields on a temporal scale. 

\subsection{Concept of symbolic transfer entropy}
We illustrate the concept of STE, which is currently more widely accepted than TE~\cite{kissler2020symbolic}, in Fig.~\ref{fig_STEmethodology}. 
%combining with symbolic dynamic filtering addresses that problem by converting the data from various fields into the ranking order patterns of the time series}.
For instance, we consider $(X_t)^{n}_{t=1}$ to be a $D$ dimensional time series of $X$ as follows, $X_t = (X_t,X_{t+1},X_{t+2},....,X_{t+(D-1)})$.  
We can consider a group $S_X = (X_1, X_2,..., X_D)$ to be $\Pi =(l_1,l_2,..,l_D)$ type symbol if and only if following conditions are satisfied:

(a) $X_{l_1}\le X_{l_2}\le X_{l_3}\le$....$\le X_{D}$,

(b) $l_{k-1}\le l_{k}$ if $X_{l_{k-1}} = X_{l_{k}}$.

Thus, based on a ranking order of the time series patterns (condition (a)) and uniqueness of the symbols (condition (b)), we can form a symmetric group of $D!$ for $D \ge 2$ when all the symbols of length $D$ will be formed. Using the concept of symbolic dynamics flitering, Eq.~(\ref{eq:TE}) can be written as~\cite{staniek2008symbolic},
\begin{equation}
    T^S_{X{\rightarrow}Y} = \sum_{n} P(\Pi_{Y(t_{n+1})},\Pi_{Y(t_n)},\Pi_{X(t_n)})log_{2}\frac{P(\Pi_{Y(t_{n+1})}|\Pi_{Y(t_n)},\Pi_{X(t_n)})}{P(\Pi_{Y(t_{n+1})}|\Pi_{Y(t_n)})},
\label{eq:STE}
\end{equation}
where, $\Pi_{Y(t_{n+1})}$ and $\Pi_{Y(t_{n})}$ are the ranking orders or symbol types of $Y$ at time steps $t_{n+1}$ and $t_{n}$ respectively. Similarly, $\Pi_{X(t_{n})}$ is the symbol type of variable $X$ at time step $t_{n}$. The joint and conditional probabilities of Eq.~(\ref{eq:STE}) represent the same as that stated in Eq.~(\ref{eq:TE}). The conditional probability in STE calculations aids us in extracting the asymmetric flow information between the variables.  
Further, introducing a time delay in calculating those probabilities helps better to understand the correlation between the variables~\cite{staniek2008symbolic}. Thus, by calculating STE between two variables, we can extract the information on the flow of influences that one variable has on the other.

Following Eq.~(\ref{eq:STE}), 
we measure the spatiotemporal evolutions of STE to both qualitatively and quantitatively understand the mutual influences between the reacting flow variables in a liquid rocket engine combustion system during TAI in Sec.~\ref{sec:results}. 
We also measure the net degree of mutual influence or directionality between two system variables $X$ and $Y$, as $\Delta S_{{X}{\rightarrow}{Y}}$ = $T^{S}_{{X}{\rightarrow}{Y}}$ - $T^{S}_{{Y}{\rightarrow}{X}}$, to identify the dominant feedback pathways during TAI. 
Thus, by estimating the directionality between two variables, we can identify the variable that has a stronger influence and plays a crucial role in sustaining the limit cycle oscillations during the state of TAI.

\begin{figure*}%
\centering
\includegraphics[width=1.02\textwidth]{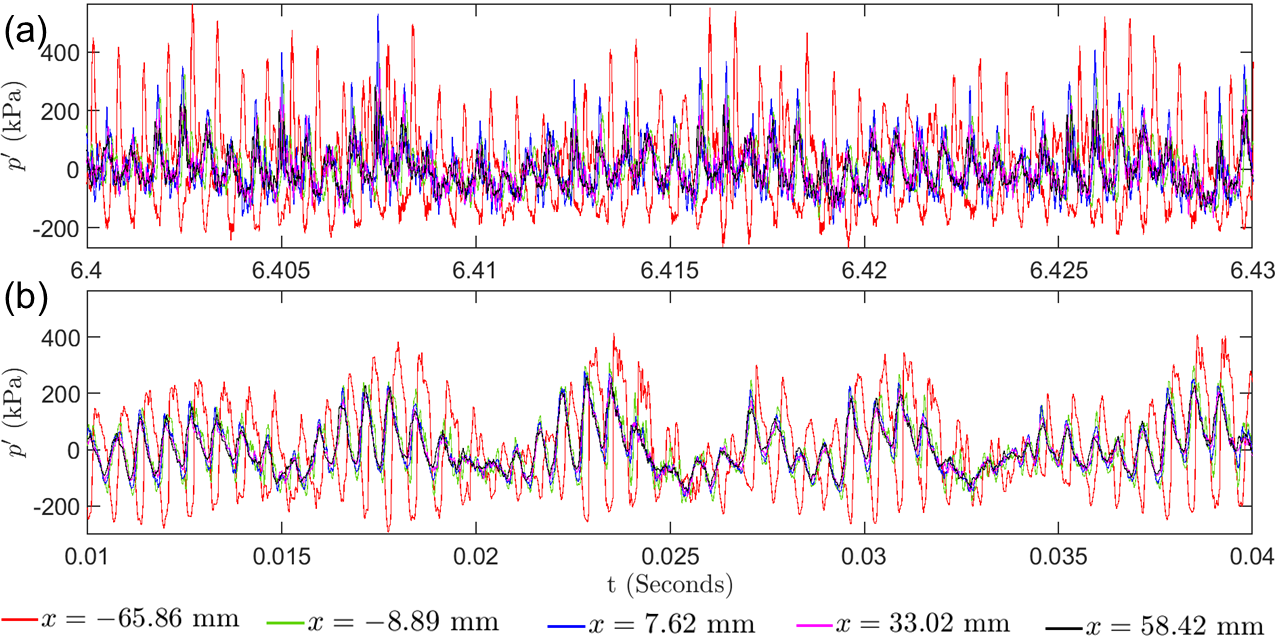}
\caption{(a) presents the fluctuations in the acoustic pressure (noted as $p^{\prime}$) at the oxidizer post ($x$ = $-65.86$ mm), the recess injector ($x$ = $-8.89$ mm) and various downstream locations of the dump plane ($x$ = 7.62 mm, 33.02 mm and 58.42 mm) obtained from experiments during TAI at the preheated oxidizer temperature of 705 K using the pressure transducers installed at those targeted locations. (b) presents the time series of $p^{\prime}$ that are obtained from CFD simulation during TAI at the same operating condition. The first 10 ms of simulation data that includes initial transients after ignition is not considered in (b). Although the lower amplitude pressure oscillations are visible in simulation data at 25 ms, the overall trend of the oscillations of simulation data is similar to that of experimental data.      
}\label{fig_timeseries}
\end{figure*}

\section{Results and Discussions}\label{sec:results}

To understand the mechanism underlying the sustenance of thermoacoustic instability (TAI) in the HAMSTER combustor, we analyze the temporal and spatial variations of the primary variables in this study. First, we discuss the time series of acoustic pressure oscillations in Sec.~\ref{subsec:time series} to understand the dynamics of self-sustained thermoacoustic oscillations at the different locations of interest. Then, we present the power spectral densities obtained using Fast Fourier transform to find the dominant modes of thermoacoustic oscillations in HAMSTER combustor. 
%in Sec.~\ref{subsec:FFT}.
Next, in Sec~\ref{subsec:STE on experimental}, we present an illustration of the concept of symbolic transfer entropy using the time series data of acoustic pressure and OH* data obtained from experiments at the wall portion that is closer to the dump region ($x$ = 7.62 mm). 
%($x$ = 0.00762 m). 
Following this, we try to understand the mutual spatiotemporal influences among the primary variables during a rise in acoustic pressure fluctuations above zero to its local maxima in Sec.~\ref{subsec:STE on spatiotemporal}. We also quantify the degree of influences between the variables using the directionality measure to determine the driving mechanism during one cycle of acoustic pressure oscillations (when approaching towards the local minima and maxima) in Sec.~\ref{subsec:STE_quantification}. 
In these subsequent sections, we use notation $()^{\prime}$ 
%(= $()-\bar{()}$, where $\bar{()}$ indicates the mean) 
to show the 
%mean subtracted 
fluctuations of the variable $()$
%, while the notation $()$ indicates the actual values of that variable 
in the discussion.   
%Finally, we summarize the significant inferences in Sec.~\ref{sec:conclusions}.    

\subsection{Time series and amplitude spectrum of acoustic pressure}\label{subsec:time series}

Figures~\ref{fig_timeseries} (a) and (b) display the temporal evolutions of acoustic pressure fluctuations ($p^{\prime}$) at the locations of interest. Based on the previous investigations~\cite{harvazinski2013analysis,fuller2019dynamic,harvazinski2020large}, the locations of interest in the present study are the oxidizer post (position of pressure transducer pt1, $x$ = $-65.86$ mm), the injector recess (position of pressure transducer pt2, $x$ = $-8.89$ mm), the head end of the combustor or closer to the dump region (position of pressure transducer pt3, $x$ = 7.62 mm), %locations of the inlet of the combustion chamber 
locations downstream of the combustor dump plane (positions of pressure transducers pt4 and pt5, where, $x$ = 33.02 mm and 58.42 mm respectively). The locations of the transducers are marked in Fig.~\ref{fig_setup}. We find from both experiment (Fig.~\ref{fig_timeseries} (a)) and CFD simulation (Fig.~\ref{fig_timeseries} (b)) that $p^{\prime}$ exhibits large amplitude
%appear more violently 
near the oxidizer post (see the red time trace) which is a consequence of the vortex shedding at the propellant separator~\cite{fuller2019dynamic} (also called as fuel collar). 
%as the signal shows more deviations about the zero value. 
%The positive feedback between the flow and acoustic fields can be the reason for the acoustic wave modulation at that upstream section. 
%In addition, observing a few aperiodic oscillations for a small time duration after successive periodic oscillations, we understand that the moving mean pressure sags due to the hydrodynamic chugging as also mentioned in the previous literature~\cite{fuller2019dynamic}.
We find almost similar temporal behavior of $p^{\prime}$ at the injector recess and the subsequent downstream locations in the combustor. 

\begin{figure*}%
\centering
\includegraphics[width=1.0\textwidth]{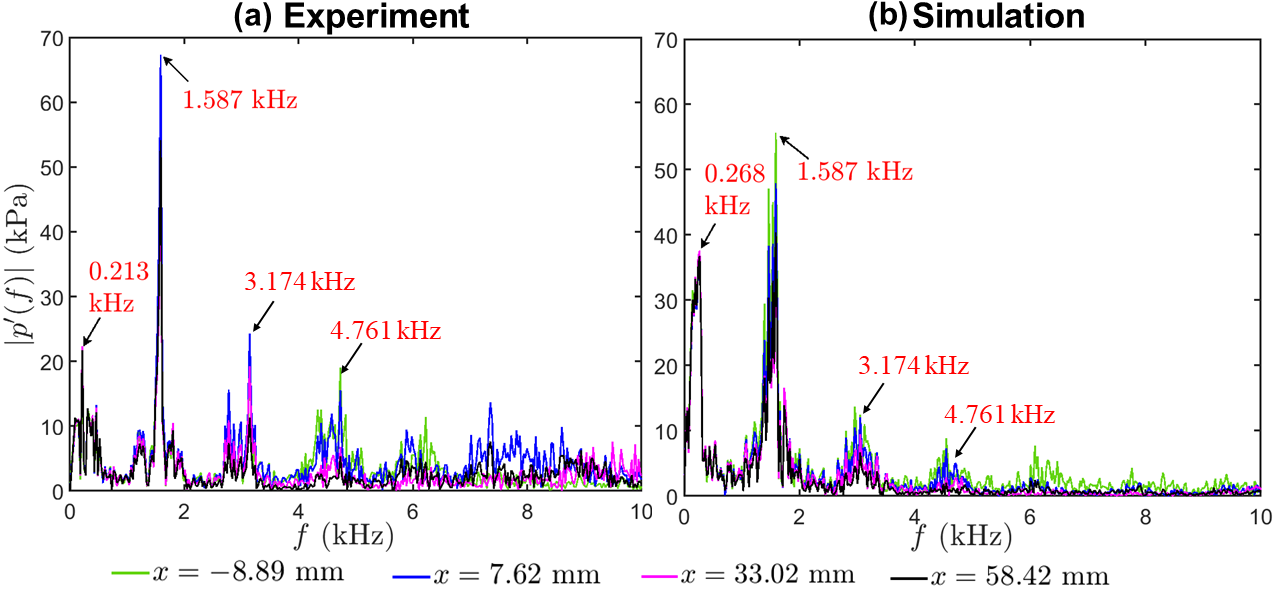}
\caption{(a) presents the FFT of time series of acoustic pressure oscillations at the injector recess ($x$ = $-8.89$ mm) and at different downstream positions of the dump plane, which are 7.62 mm, 33.02 mm, and 58.42 mm, respectively. As pointed out by the previous literature~\cite{harvazinski2020large}, the first longitudinal mode (1L) dominates at 1587 Hz. Besides, we notice another mode at 213 Hz from experimental data. From an earlier experimental study on the same combustor~\cite{fuller2019dynamic}, a bulk mode instability at a low-frequency mode below 260 Hz was observed for preheated oxidizer at 700 K and $\Phi$ = 0.8. Hence, the 213 Hz tone is attributed to the same bulk mode. (b) shows the FFT of the time series of acoustic pressure oscillations obtained from CFD simulation data at the same locations said in (a). 
%We again confirm the value of the dominant frequency of longitudinal oscillations from the simulation data here. 
The dominant longitudinal frequency mode obtained from CFD simulation data exactly matches that obtained from the experiments. However, based on CFD simulation data, the range of bulk mode frequency varies between 144 Hz and 268 Hz.}\label{fig_frequency compare}
\end{figure*}
 
Next, we perform a Fast Fourier transform (FFT) analysis to find out the dominant modes of frequencies of thermoacoustic oscillations in HAMSTER combustor using data obtained from both experiments and CFD acoustic pressure data in Figs.~\ref{fig_frequency compare} (a) and (b), respectively. We observe the presence of two modes, one having a low frequency ($<$ 300 Hz), and another is seen at 1587 Hz~\cite{fuller2019dynamic}. Previous works~\cite{fuller2019dynamic,harvazinski2020large} performed by the Purdue group mentioned that the low frequency mode is a hydrodynamic or bulk mode instability resulting from the fluctuations in the propellant feed system that can be brought on by abrupt changes in the fluid properties within the system, and a delayed combustion response. 
%Such low frequency mode instabilities are regarded as bulk mode instability in rocket engines. 
However, the dominant mode is the first longitudinal mode (1L) at 1587 Hz at 705 K preheated oxidizer temperature. Previous investigation~\cite{fuller2019dynamic} found that the effect of bulk instability becomes less pronounced as the oxidizer temperature increases. At higher preheated oxidizer temperatures, $p^{\prime}$ rises and consequently, the amplitude of thermoacoustic mode becomes higher than that of bulk instability. However, although one mode between the bulk and the longitudinal (1L) modes often dominates in liquid rocket engine combustors, both modes coexist. The system may temporarily and intermittently switch between them that confirms the interdependence between those low and high frequency modes, as demonstrated in the previous literature~\cite{fuller2019dynamic} at 700 K preheated oxidizer temperature.
On the other hand, the harmonics of 1L mode are observed (2L - 3174 Hz and 3L - 4761 Hz, see Fig.~\ref{fig_frequency compare}).
%~\cite{fuller2019dynamic}. 
The amplitudes of the 
%longitudinal modes 
2L and 3L modes are found to be relatively low in the amplitude spectrum of the CFD data compared to the experimental data. 
%Further, from a bispectral analysis at 700 K preheated oxidizer temperature, it was found that the low (bulk) and high (1L) frequency modes are interdependent~\cite{fuller2019dynamic}. 
%\textcolor{blue}{However, although one mode between bulk and longitudinal (1L) often dominates in liquid rocket engine combustors, both modes coexist}. The system may temporarily and intermittently switch between them \textcolor{blue}{that confirms the interdependence between those low and high frequency modes, as demonstrated in the previous literature~\cite{fuller2019dynamic} at 700 K preheated oxidizer temperature.}  

\begin{figure*}[ht!]%
\centering
\includegraphics[width=1.0\textwidth]{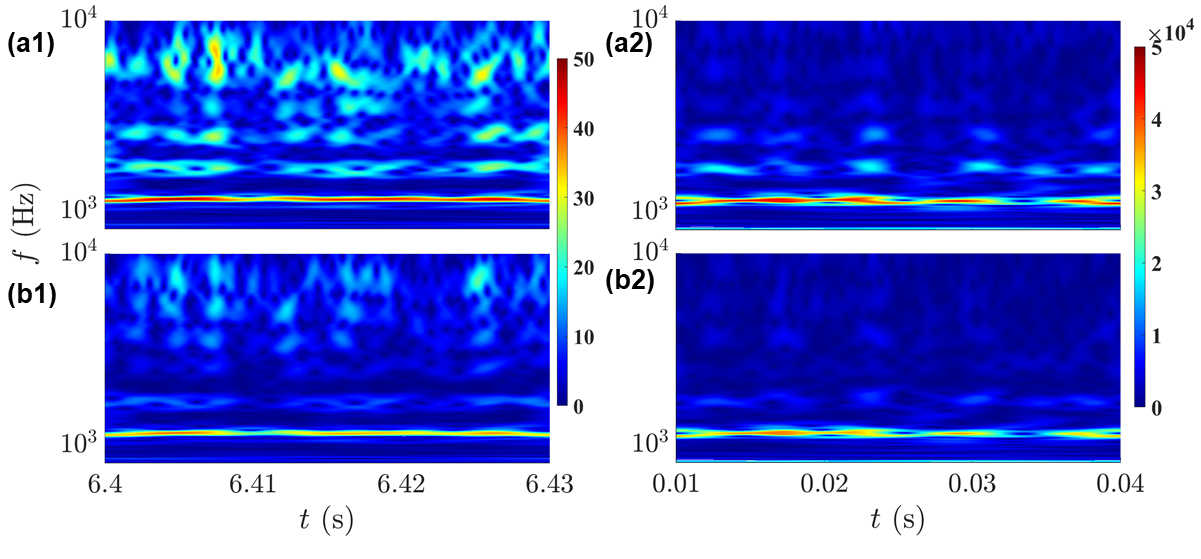}
\caption{(a1-b1) and (a2-b2)  show the scalograms (frequency vs time) of experimentally and numerically (CFD) obtained $p^{\prime}$ during combustion instability at two locations of the HAMSTER combustion chamber: 7.62 mm (a1 $\&$ a2) and 58.42 mm (b1 $\&$ b2) downstream from the dump plane or head end of the combustor. A dominant 1L mode is clearly observed during the oscillations near the dump plane region. Due to the presence of a pressure anti-node at 7.62 mm, the amplitude of $p^{\prime}$ (both positive and negative) becomes very high. Therefore, the amplitude of 1L mode becomes high at that location in the combustion chamber. However, the amplitude of the 1L mode gradually becomes low downstream of the combustion chamber.}\label{fig_scalogram}
\end{figure*}

Further, in Figs.~\ref{fig_scalogram} (a1)-(b1) and (a2)-(b2), we try to observe the temporal evaluations of the power spectrum in the downstream direction of the combustor dump plane. We find that 1L mode is strong closer to the dump region or the head end of the combustor ($x$ = 7.62 mm) throughout the time period of 30 ms (Figs.~\ref{fig_scalogram} (a1) and (a2)) since
pressure anti-nodes for acoustic standing waves are noticed there. However, 1L mode becomes weaker at downstream locations %combustion chamber 
dump plane (at $x$ = 58.42 mm) (Figs.~\ref{fig_scalogram} (a2) and (b2)) and almost diminishes at $x$ = 180.34 mm (see Sec.~A in Appendix) since pressure nodes for this 1L mode are situated near $x$ = 180.34 mm. Therefore, we understand that the near dump region in the combustion chamber can be the zone of interest towards understanding the mechanism of thermoacoustic oscillations. However, the bulk modes do not diminish at the downstream locations of the combustor (Figs.~\ref{fig_scalogram}(a2)-(b2)). 

In the next section (Sec.~\ref{subsec:STE on experimental}), we try to understand the mutual influence between acoustic pressure and OH$^*$ chemiluminescence oscillations based on the temporal data found experimentally near the dump plane of the HAMSTER combustor (location of pressure transducer pt3).                       

\begin{figure*}%
\centering
\includegraphics[width=0.745\textwidth]{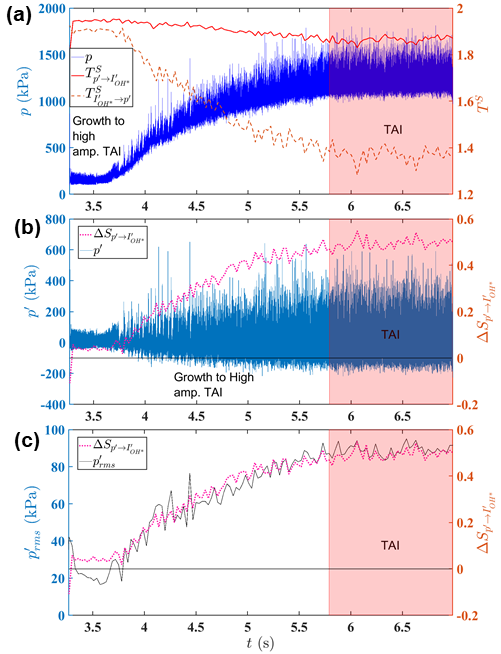}
\caption{The amplitude of acoustic pressure (i.e., $p = \overline{p}+p^{\prime}$) measured from experiments in HAMSTER and its mean subtracted local fluctuations (noted as $p^{\prime}$) are presented on the left vertical axis in the plots (a) and (b) respectively. The variation of symbolic transfer entropy based on $p^{\prime}$ and $I^{\prime}_{OH^*}$ obtained at $x$ = 7.62 mm near the combustor wall is shown on the right vertical axis. %Note that OH* approximates the heat release rate (HRR) for the experimental data as direct measurement of HRR is not possible. 
We consider the time duration when a growth of low amplitude oscillations saturates to high amplitude limit cycle oscillations. 
%Further, note that the low amplitude periodic oscillations are the chug instabilities. 
During the growth when the bulk mode and the 1L mode oscillations co-exist, we find that the influence of $I^{\prime}_{OH^*}$ on $p^{\prime}$ gradually reduces where as the influence of $p^{\prime}$ on $I^{\prime}_{OH^*}$ remains nearly constant (a). 
%To quantitatively understand the net influence of $p^{\prime}$ on $I^{\prime}_{OH^*}$, we calculate the directionality index between $p^{\prime}$ and $I^{\prime}_{OH^*}$ (i.e., $\Delta S_{{p^{\prime}}{\rightarrow}{I^{\prime}_{OH^*}}}$) in plot (b). 
We find a drop in directionality ($\Delta S_{{p^{\prime}}{\rightarrow}{I^{\prime}_{OH^*}}}$) in (b) due to the continuous reduction in the influence of $I^{\prime}_{OH^*}$ on $p^{\prime}$. The root mean square of $p^{\prime}$ is shown in (c).}
\label{fig_STE_exp}
\end{figure*}

\subsection{STE analysis on temporal data of HAMSTER combustor}\label{subsec:STE on experimental}

Based on the prior research~\cite{kasthuri2019dynamical, shima2021formation,godavarthi2018coupled,hashimoto2019spatiotemporal}, we use the temporal evolution of acoustic pressure fluctuations $p^{\prime}$ and OH$^{*}$ chemiluminescence fluctuations $I^{\prime}_{OH^*}$ of HAMSTER combustor close to the wall at $x$ = 7.62 mm to gain a basic understanding of the mechanism underlying the mutual influence between these two variables.  
%acoustic oscillations ($p^{\prime}$ ) and OH$^*$ chemiluminescence oscillations ($I^{\prime}_{OH^*}$), which also represents the heat release rate oscillations ($\dot{Q}^{\prime}$). 
 We show $p^{\prime}$ of the combustor obtained from the experiment when the propellants are throttled to the steady operation condition to obtain the state of thermoacoustic oscillations following the ignition of the main chamber (engine start) at 3.25 s in Fig.~\ref{fig_STE_exp} (b). Subsequently, $p^{\prime}$ grows from low to high amplitudes. During this growth in amplitude, we observe from earlier literature~\cite{fuller2019dynamic} that both the bulk mode and 1L mode oscillations co-exist. However, the 1L mode primarily dominates from 5.8 to 6.9 s (time series and FFT of %the 
 %acoustic pressure fluctuations
 $p^{\prime}$ are already shown in Figs.~\ref{fig_timeseries}a and~\ref{fig_frequency compare}a during 6.4 to 6.43 s), when the combustion oscillations gain sufficient acoustic power to exhibit saturated high amplitude limit cycle oscillations (HALCO). In Fig.~\ref{fig_STE_exp} (a), we estimate the STE ($T^{S}$) between %acoustic pressure oscillations (
 $p^{\prime}$
 %)  
 and $I^{\prime}_{OH^*}$ to quantitatively understand their mutual transfer of information. Note that STE calculated using the fluctuations of the variables (i.e., $p^{\prime}$ and $I^{\prime}_{OH^*}$) and actual values of the variables (i.e., $p$ and $I_{OH^*}$) are the same as the method relies on ranking order of the patterns of time series. To determine $T^{S}$ at the time instants, we employ moving overlapping windows, each with 0.03 s  data (i.e., 30000 data points that correspond to around 48 cycles during TAI). 
 %\textcolor{blue}{We find that around 48 cycles corresponds to that data size are sufficient to capture the proper dynamics of the system.} 
 In the calculation of $T^{S}$, we consider the embedding dimension ($D$) of the high dimensional system as 7, as obtained from the method of false nearest neighbors (FNN) ~\cite{abarbanel1993analysis,kennel1992determining} on both experimental and CFD acoustic pressure data. 
%\textcolor{red}{The process for finding $D$ is shown in the supplementary material.} 
%The time lag ($l$) is taken as 1.   

From the estimation of $T^{S}$ in Fig.~\ref{fig_STE_exp} (a), we find that the influence of $p^{\prime}$ on $I^{\prime}_{OH^*}$ ($T^{S}_{{p^{\prime}}{\rightarrow}{I^{\prime}_{OH^*}}}$) increases when low amplitude pressure oscillations are seen after the engine starts up. However, during the growth to saturated HALCO, 
%$p^{\prime}$ 
$T^{S}_{{p^{\prime}}{\rightarrow}{I^{\prime}_{OH^*}}}$ consistently has higher value over $T^{S}_{{I^{\prime}_{OH^*}}{\rightarrow}{p^{\prime}}}$. This signifies that the influence of $p^{\prime}$ on $I^{\prime}_{OH^*}$ ($T^{S}_{{p^{\prime}}{\rightarrow}{I^{\prime}_{OH^*}}}$) is seen to be higher than that of $I^{\prime}_{OH^*}$ on $p^{\prime}$ ($T^{S}_{{I^{\prime}_{OH^*}}{\rightarrow}{p^{\prime}}}$) at the wall near dump plane during this phase. This observation corroborates with the assertion made in earlier works from Gotoda's group~\cite{hashimoto2019spatiotemporal,shima2021formation,mori2022nonlinear,mori2023feedback}. On the other hand, $T^{S}_{{I^{\prime}_{OH^*}}{\rightarrow}{p^{\prime}}}$ significantly reduces until HALCO is observed. Hence, the feedback interaction from heat release rate ($\propto$ OH* chemiluminescence)  to acoustic pressure fluctuations plays a vital role during the growth of low amplitude oscillations to HALCO.

In Figs.~\ref{fig_STE_exp} (b)-(c), we estimate the directionality index between $p^{\prime}$ and $I^{\prime}_{OH^*}$ (i.e., $\Delta S_{{p^{\prime}}{\rightarrow}{I^{\prime}_{OH^*}}}$ = $T^{S}_{{p^{\prime}}{\rightarrow}{I^{\prime}_{OH^*}}} - T^{S}_{{I^{\prime}_{OH^*}}{\rightarrow}{p^{\prime}}}$). We find that $\Delta S_{{p^{\prime}}{\rightarrow}{I^{\prime}_{OH^*}}}$ 
%becomes above 
acquires a value greater than zero at $t$ = 3.3 s 
%after the engine starts up 
due to an asymmetric
bidirectional coupling between $p^{\prime}$ and the heat release rate with a stronger
influence of $p^{\prime}$ on the heat release rate than vice versa
%$T^{S}_{{p^{\prime}}{\rightarrow}{I^{\prime}_{OH^*}}}$ dominates over $T^{S}_{{I^{\prime}_{OH^*}}{\rightarrow}{p^{\prime}}}$)
before $p^{\prime}_{rms}$ (see in Fig.~\ref{fig_STE_exp} (c)) starts to increase. After that, $\Delta S_{{p^{\prime}}{\rightarrow}{I^{\prime}_{OH^*}}}$ gradually increases during the growth to HALCO. \textcolor{blue}{}Earlier studies~\cite{hashimoto2019spatiotemporal,shima2021formation,mori2022nonlinear,mori2023feedback} reported that the direction of the feedback interaction starts to dominate from acoustic pressure fluctuations to the heat release rate fluctuations in combustor when TAI is formed with strong periodicity. We unveil that $\Delta S_{{p^{\prime}}{\rightarrow}{I^{\prime}_{OH^*}}}$ increases close to the wall at $x=7.62$ mm  during a transition to HALCO as $T^{S}_{{I^{\prime}_{OH^*}}{\rightarrow}{p^{\prime}}}$ decreases, while a nearly constant $T^{S}_{{p^{\prime}}{\rightarrow}{I^{\prime}_{OH^*}}}$ is observed. 

We further compare the result with that obtained using CFD data (see Table~\ref{Tab:3}) for verification. We calculate temporally averaged directionality index between ${p^{\prime}}$ and the intensity of $OH$ chemiluminescence emission from both experimental and CFD data at the three locations of the pressure transducers (pt3, pt4 and pt5) in the combustor during fully developed TAI. The estimation of the index is done over a time duration of 30 ms as the CFD simulations are limited to about 30 ms of run time due to the computational cost. We find the positive sign of temporally averaged directionality index between ${p^{\prime}}$ and the intensity of $OH$ chemiluminescence emission for both experimental and CFD data at the prescribed locations during fully developed TAI. It is important to note that the predominant direction of feedback between these two variables is found from acoustic pressure to $OH$ chemiluminescence emission using both experimental and simulation databases, although the calculated values are not exactly the same.     
%In this way, based on the temporal variations of the variables, we can find the reason for the dominance of a variable and  identify the direction of dominant feedback interaction between them.} 

\begin{table*} % Table 4
\caption{We calculate $\Delta S_{{p^{\prime}}{\rightarrow}{I^{\prime}_{OH*}}}$ (from experimental data using photo multiplier tube) or $\Delta S_{{p^{\prime}}{\rightarrow}{I^{\prime}_{OH}}}$ (from CFD simulation data) over a time duration of 30 ms. Note that the only difference between $OH^{*}$ and $OH$ is that $OH^{*}$ is the line of sight integrated measure using photo multiplier tube during experiment close to the wall at $x$ =7.62 mm and $OH$ is measured from the computational domain using CFD at the same location. We check the sign convention of the directionality measure (> 0 or < 0) during TAI at the following locations: (pt3) position of 1$^{st}$ pressure transducer downstream of dump plane (result is shown in Fig.~\ref{fig_STE_exp}) : $x = 7.62$ mm, $10$ mm $\le y \le12.5$ mm, (pt4) position of 2$^{nd}$ pressure transducer downstream of dump plane: $x = 33.02$ mm, $10$ mm $\le y \le12.5$ mm, (pt5) position of 3$^{rd}$ pressure transducer downstream of dump plane: $x = 58.42$ mm, $10$ mm $\le y \le12.5$ mm.}\label{Tab:3}
\centering
%m{'width'}
\setlength\arrayrulewidth{1.5 pt}
%\begin{tabular}{|p{2.3cm}|p{2.3cm}|p{2.3cm}|p{2.3cm}|p{2.3cm}|p{2.3cm}|}
%\begin{tabular}{|p{2.3cm}|p{3.1 cm}|p{2.5cm}|p{4cm}|p{3cm}|p{2.5cm}|}
\begin{tabular}{p{2cm}|p{4cm}|p{4cm}|p{4cm}}
%\begin{tabular}{ |l|l|l|l|l|l| }
\hline
\begin{flushleft}Serial No \end{flushleft}&\begin{flushleft}Location\end{flushleft} & \begin{flushleft}Data type\end{flushleft} & \begin{flushleft} temporally averaged $\Delta S_{{p^{\prime}}{\rightarrow}{I^{\prime}_{OH^*}}}$ or $\Delta S_{{p^{\prime}}{\rightarrow}{I^{\prime}_{OH}}}$ \end{flushleft}\\
\hline
1 & Location pt3  & Experimental & 0.462 \\
%\cline{2-4}
2 & Location pt4  & Experimental & 0.357\\
%\cline{2-6}
3 & Location pt5  & Experimental & 0.665\\
4 & Location pt3  & CFD & 0.028 \\
%\cline{2-4}
5 & Location pt4  & CFD & 0.046\\
%\cline{2-6}
6 & Location pt5 & CFD & 0.048\\
\hline
\end{tabular}
%\
\end{table*}

\begin{figure*}%
\centering
\includegraphics[width=0.9\textwidth]{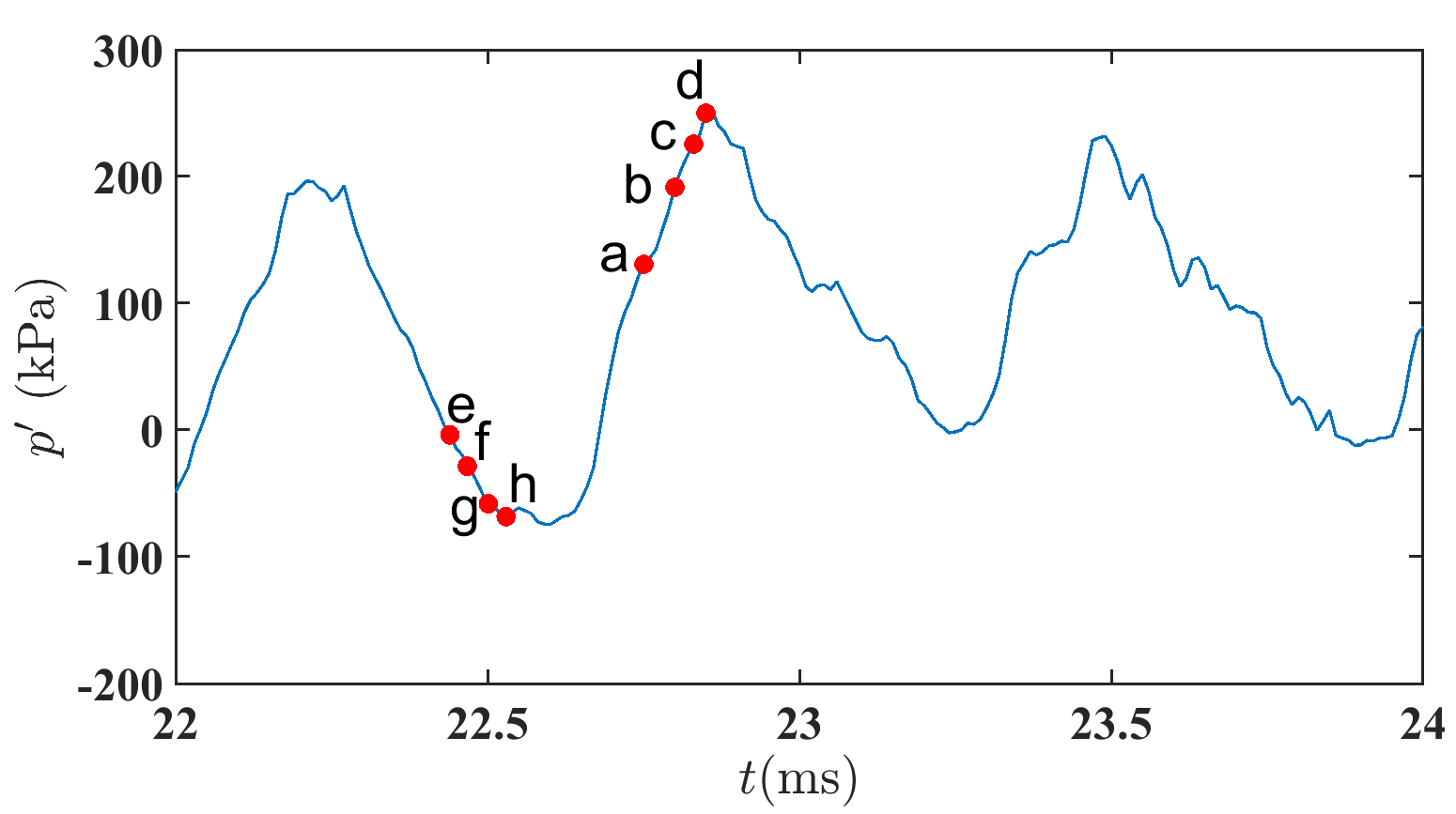}
\caption{We consider four time instants when the acoustic pressure approaches a local maxima during thermoacoustic oscillations for the spatiotemporal analysis: 22.75 ms (a), 22.80 ms (b), 22.83 ms (c) and 22.85 ms (d). Also, we take four time instants (e-h) for understanding the mutual interaction between $p^{\prime}$ and $\dot{Q}^{\prime}$ when acoustic pressure approaches a local minima in Appendix B: 22.44 ms (e), 22.47 ms (f), 22.50 ms (g) and 22.53 ms (h).}
\label{fig_timeseriesmarked}
\end{figure*}

Nevertheless, to understand the mechanism or evolution of the complex states in a system, a conclusion drawn solely from temporal data may provide only partial information while the understanding on spatiotemporal scale can offer a deeper inference about a system's dynamics based on which one can identify the "critical region"~\cite{krishnan2021suppression,raghunathan2022seeds} or epicenter of TAI at which the bidirectional coupled interactions between the variables are seen to be more promising during the sustenance the combustion instability in the liquid rocket engine combustors. Therefore, in the subsequent sections (Secs.~\ref{subsec:STE on spatiotemporal} and~\ref{subsec:STE_quantification}), we try to understand the mechanism associated with TAI through the direction of mutual information between the primary variables of the HAMSTER combustor on spatiotemporal scale.

\subsection{STE analysis on spatiotemporal behavior of HAMSTER combustor}\label{subsec:STE on spatiotemporal}

\begin{sidewaysfigure*}%
\centering
\includegraphics[width=0.9\textwidth]{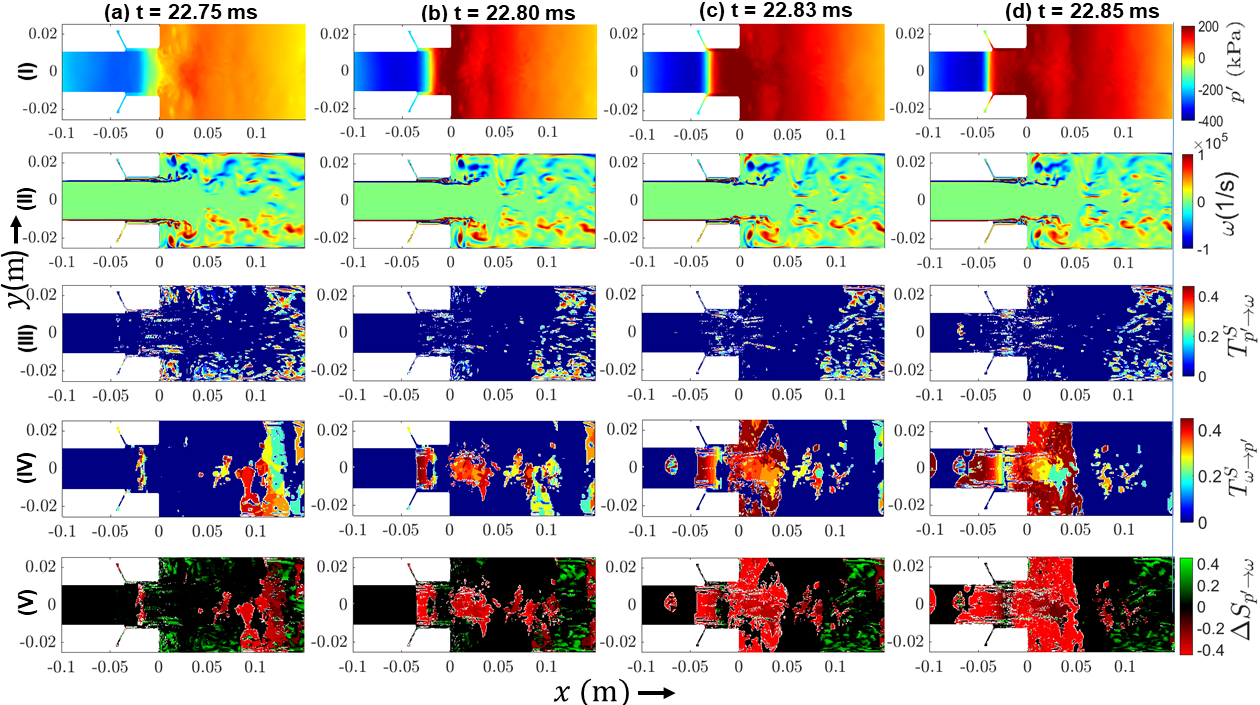}
\caption{$p^{\prime}$, $\omega$, STE from $p^{\prime}$ to $\omega$ ($T^{S}_{{p^{\prime}}{\rightarrow}{\omega}}$), STE from $\omega$ to $p^{\prime}$ ($T^{S}_{{\omega}{\rightarrow}{p^{\prime}}}$) and the directionality ($\Delta S_{{p^{\prime}}{\rightarrow}{\omega}}$)  are shown (from top to bottom) at (a) $t$ = 22.75 ms, (b) $t$ = 22.80 ms, (c) $t$ = 22.83 ms and (d) $t$ = 22.85 ms. We find that influence of $\omega$ on $p^{\prime} ($ $T^{S}_{{\omega}{\rightarrow}{p^{\prime}}}$) spreads near the dump plane and recess injector when the acoustic pressure amplifies spatially at the dump plane during $t=22.75-22.85$ ms. Further, from $\Delta S_{{p^{\prime}}{\rightarrow}{\omega}}$, we identify the rapid emergence of the relatively stronger influencing regions of $\omega$ over $p^{\prime}$ near the dump plane. The dimension of the combustor is shown in meter.}
\label{fig_STE_simu_p_omega_acoustic rising}
\end{sidewaysfigure*}

In the previous section (Sec.~\ref{subsec:STE on experimental}), we use the acoustic pressure and heat release rate (chemiluminescence) fluctuations obtained experimentally during the growth of pressure oscillations to TAI. However, in this section, we focus on the acoustic, %hydrodynamic 
flow and combustion variables obtained using CFD simulation on spatiotemporal scales during TAI. We take two 
%hydrodynamics 
flow variables: vorticity and velocity. Besides, two combustion variables, the heat release rate and the temperature of combustion field, are considered in the study. First, we describe the observations obtained by analyzing the spatiotemporal fluctuations of these variables. Following this, we obtain a qualitative understanding of the direction of mutual interaction between these variables 
by performing  an analysis based on the concept of STE on their oscillations. 
%(denoted by  $A^{\prime}$, where, $A$ represents the variable) near the dump plane (i.e., head end) of the HAMSTER combustor during TAI.
Hereon, %the 
%actual values 
vorticity and the fluctuations in 
%acoustic pressure, 
heat release rate, flow velocity and temperature are denoted by the notations $\omega$, 
%$p^{\prime}$, 
$\dot{Q}^{\prime}$, $u^{\prime}$, and $\theta^{\prime}$ respectively in the discussion of their spatiotemporal evolutions. For the spatiotemporal analysis of the variables (Figs.~\ref{fig_STE_simu_pq_acoustic rising} to~\ref{fig_STE_simu_p_temp_acoustic rising}), we consider the time stamps when $p^{\prime}$ significantly grows and approaches the local maxima during $t = 22.75$ to $22.85$ ms as marked in Fig~\ref{fig_timeseriesmarked}. Note that this time scale is different than the experiment, and this period is obtained after a steady state has been achieved. However, the inlet conditions and operating parameters are all kept exactly the same in the simulation as in the experiments, as mentioned previously, to replicate similar dynamics during TAI.     

%\textcolor{red}{DONE}

\subsubsection{ Mutual influence between the acoustic and heat release rate fields}\label{subsubsec:acoustic-heat}

%\onecolumngrid
%\begin{center}
\begin{sidewaysfigure*}
\centering
%\hspace{5cm}
\includegraphics[scale=0.65]{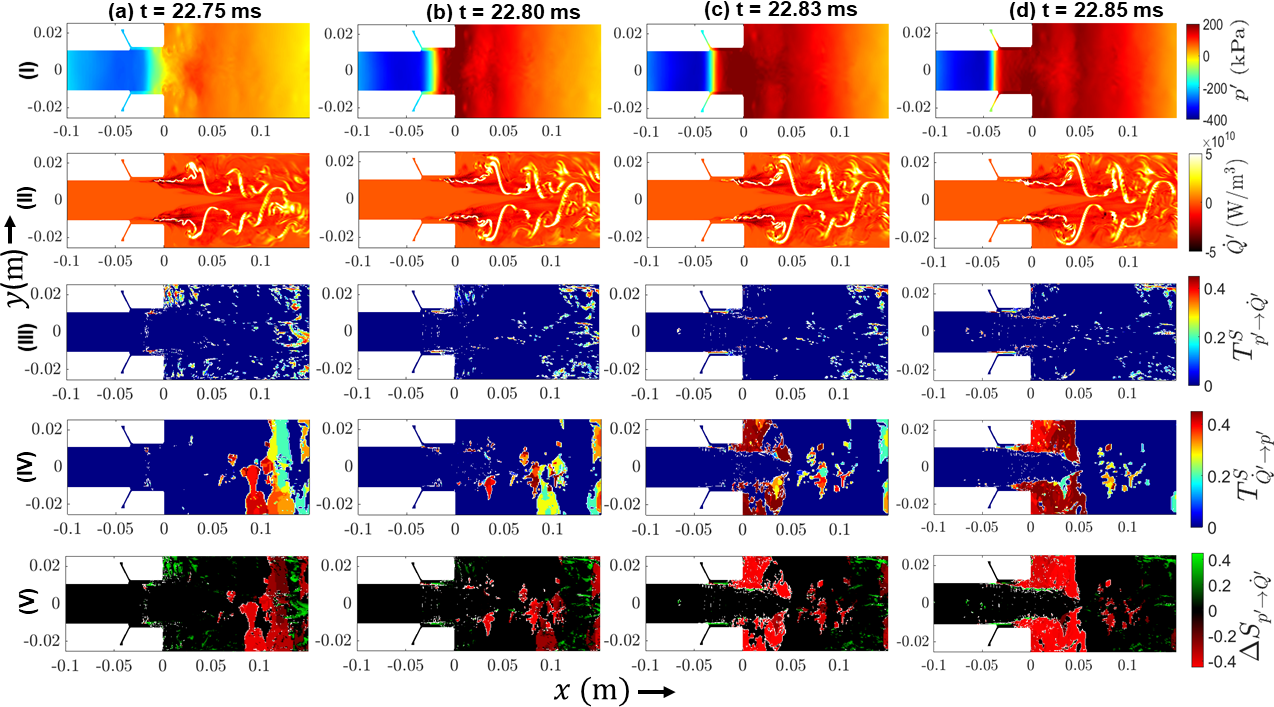}
\caption{$p^{\prime}$ (row-I) and $\dot{Q}^{\prime}$ (row-II) are shown at different time instants when $p^{\prime}$ grows rapidly near the dump plane area ($-0.1$ m $\le x \le 0.15$ m, $z=0$, and $-0.025$ m $\le y \le 0.025$ m) between $t = 22.75$ ms and $t = 22.85$ ms. Simultaneously, STE from $p^{\prime}$ to $\dot{Q}^{\prime}$ ($T^{S}_{{p^{\prime}}{\rightarrow}{\dot{Q}^{\prime}}}$), STE from $\dot{Q}^{\prime}$ to $p^{\prime}$ ($T^{S}_{{\dot{Q}^{\prime}}{\rightarrow}{p^{\prime}}}$) and the directionality index ($\Delta S_{{p^{\prime}}{\rightarrow}{\dot{Q}^{\prime}}}$) (from top to bottom) are analyzed at (a) $t$ = 22.75 ms, (b) $t$ = 22.80 ms, (c) $t$ = 22.83 ms and (d) $t$ = 22.85 ms. From the STE analysis, we find ${p^{\prime}}{\rightarrow}{\dot{Q}^{\prime}}$ influence near the wall at the dump region (see $T^{S}_{{p^{\prime}}{\rightarrow}{\dot{Q}^{\prime}}}$ at III (a)). However, the influence ${\dot{Q}^{\prime}}{\rightarrow}{p^{\prime}}$ becomes significant when the acoustic field emerges near the dump plane during (b)-(d) (see $T^{S}_{{\dot{Q}^{\prime}}{\rightarrow}{p^{\prime}}}$). The examination of $\Delta S_{{p^{\prime}}{\rightarrow}{\dot{Q}^{\prime}}}$ allows us to precisely pinpoint the predominant direction between $p^{\prime}$ and $\dot{Q}^{\prime}$. We detect  the switching in the direction of predominant influence from $p^{\prime}{\rightarrow}\dot{Q}^{\prime}$ (green regions) into $\dot{Q}^{\prime}{\rightarrow}p^{\prime}$ (red colored area)  at the wall near the dump plane (V[(b)-(d)]) 
%of $\Delta S_{{p^{\prime}}{\rightarrow}{Q^{\prime}}}$)
. Note that the dimension of the combustor is indicated in meter (m).}\label{fig_STE_simu_pq_acoustic rising}
%\label{fig_STE_simu_pq_acoustic rising}
\end{sidewaysfigure*}
%\end{center}
%\twocolumngrid

%\textcolor{red}{I feel all the figure sizes from now on can be increased}
In Fig.~\ref{fig_STE_simu_pq_acoustic rising}, we present the spatial distributions of $p^{\prime}$ and $\dot{Q}^{\prime}$ (considering $-0.1$ m $\le x \le$ 0.15 m, $z=0$ and $-0.025$ m $\le y \le$ 0.025 m)  
during $t = 22.75$ to $22.85$ ms near the dump region. As $p^{\prime}$ rises, $\dot{Q^{\prime}}$ also rises in the shear layer and is seen to be higher from the wall to the centerline. The consumption of large amount of fuel and the movement of the burning fuel trapped in the recirculation zone (near the dump plane) to the centerline coincide with a rise in overall heat release rate fluctuations as mentioned in the previous study~\cite{harvazinski2020large}. In addition, 
%the amount of heat release rate 
$\dot{Q}^{\prime}$ increases in the vicinity of the fuel collar where flame stabilizes during the period. These observations raise the following questions:

1) Does the increase in $\dot{Q}^{\prime}$ along the shear layer influence rise in $p^{\prime}$? 

2) Does $p^{\prime}$ have influence on $\dot{Q}^{\prime}$ at the fuel collar?

In this section, we try to answer the questions through the variation in mutual influence between $p^{\prime}$ and $\dot{Q}^{\prime}$ using STE method (III to IV rows of Fig.~\ref{fig_STE_simu_pq_acoustic rising}).  

From the analysis, at $t=22.75$ ms, we find a %significant 
%few number of cells 
few pockets having high $T^{S}_{{p^{\prime}}{\rightarrow}{\dot{Q}^{\prime}}}$ near the wall of the dump region and at the downstream direction of the dump plane (see $T^{S}_{{p^{\prime}}{\rightarrow}{\dot{Q}^{\prime}}}$ in Fig.~\ref{fig_STE_simu_pq_acoustic rising} III (a)), which signifies that $p^{\prime}$ influences $\dot{Q}^{\prime}$  
%the combustion field 
at those zones. The influence from $p^{\prime}$ induces the fluctuations in heat release rate field, causing the wrinkling in the flame. 
%However, the influence $p^{\prime}{\rightarrow}Q^{\prime}$ is asymmetric at $t=22.75$ ms. 
On the other hand, the influence of $p^{\prime}{\rightarrow}\dot{Q}^{\prime}$ almost diminishes at the wall of the dump regime during $t=22.80$ ms to $22.85$ ms (see $T^{S}_{{p^{\prime}}{\rightarrow}{\dot{Q}^{\prime}}}$ in Fig.~\ref{fig_STE_simu_pq_acoustic rising} III [(b)-(d)] ), while a gradually thickening of the line of $T^{S}_{{p^{\prime}}{\rightarrow}{\dot{Q}^{\prime}}}$ in the vicinity of fuel collar indicates an influence of $p^{\prime}$ on $\dot{Q}^{\prime}$. 
%Note that $T^{S}_{{p}{\rightarrow}{\dot{Q}}}$ and $T^{S}_{{p^{\prime}}{\rightarrow}{\dot{Q}^{\prime}}}$ are essentially same since STE method relies on the ranking of the patterns of the variables.  

In contrast to the spatial distributions of $T^{S}_{{p^{\prime}}{\rightarrow}{\dot{Q}^{\prime}}}$, a rapid 
%increment in the number of cells 
expansion of the zones having higher $T^{S}_{{\dot{Q}^{\prime}}{\rightarrow}{p^{\prime}}}$ near dump regime after $t=22.80$ ms indicates the increment of influence of $\dot{Q}^{\prime}$ on $p^{\prime}$ (see $T^{S}_{{\dot{Q}^{\prime}}{\rightarrow}{p^{\prime}}}$ in Fig.~\ref{fig_STE_simu_pq_acoustic rising} IV[(b)-(d)]). In other words, the increase in $T^{S}_{{\dot{Q}^{\prime}}{\rightarrow}{p^{\prime}}}$  indicates the evolution of the interaction zones for heat release rate where it significantly influences the acoustic field. Interestingly, the influence of $\dot{Q}^{\prime}{\rightarrow}p^{\prime}$ becomes almost symmetric about the centerline after $t=22.80$ ms, signifying that the influence from $\dot{Q}^{\prime}$ to $p^{\prime}$ occurs throughout the dump plane. Godavarthi et al.~\cite{godavarthi2018coupled} also found the predominance of $\dot{Q}^{\prime}$ over $p^{\prime}$ during a phase synchronization state (where phases of $p^{\prime}$ and $\dot{Q}^{\prime}$ appear to be locked in time) and a general synchronization state (where both amplitudes and phases of $p^{\prime}$ and $\dot{Q}^{\prime}$ appear to be locked in time) in a bluff body stabilized gas turbine combustor based on the temporal data.     

Further, we can identify the change of the dominant direction of the feedback interaction between $p^{\prime}$ and $\dot{Q}^{\prime}$ in both space and time from the analysis of $\Delta S_{{p^{\prime}}{\rightarrow}{\dot{Q}^{\prime}}}$ (see Fig.~\ref{fig_STE_simu_pq_acoustic rising} V). Comparing the mutual influences between $p^{\prime}$ and $\dot{Q}^{\prime}$ during $t$ = 22.75 to 22.85 ms, we detect a sharp transition in the directionality between these two variables by observing the switching of the regime where influence of $p^{\prime}$ on $\dot{Q}^{\prime}$ is dominant (green colored regions having positive values) into  the regime where influence of $\dot{Q}^{\prime}$ on $p^{\prime}$ is dominant (red colored regions having negative values). The change in directionality from $p^{\prime}$ to $\dot{Q}^{\prime}$ signifies that the influence of $\dot{Q}^{\prime}$ on $p^{\prime}$ is stronger than that of $p^{\prime}$ on $\dot{Q}^{\prime}$. Interestingly, along with the dump region, the top of the fuel collar or fuel-oxidizer splitter plate also shows the same shift, indicating that the flame produced from the fuel side plays a crucial role in the oscillatory behavior
of the acoustic field. 
%that is switched from $p^{\prime}$ (green regions) to $\dot{Q}{\prime}$ (red colored regions) at the region close to the wall of the dump plane. 
%A strong influence of $Q^{\prime}$ on the $p^{\prime}$ can be further clarified from the directionality index when $p$ field oscillates at high amplitude (Figs.~\ref{fig_STE_simu_pq_acoustic rising} I and~\ref{fig_STE_simu_pq_acoustic rising} V [(b)-(d)]). 
Thus, we obtain a clear perception about the direction and strength of mutual interaction between acoustic and heat release rate fields from the information of their spatiotemporal behavior during a local maxima of thermoacoustic oscillations. 
We also try to understand the mutual influence between these two variables when $p^{\prime}$ becomes locally minimum (corresponding to Fig.~\ref{fig_timeseriesmarked}) during TAI (see Sec.~B in Appendix). Significantly, we find an emergence of the influence in the direction of $\dot{Q}^{\prime}$ to $p^{\prime}$ during this period (see Fig.~\ref{fig_STE_simu_pq_acoustic damping} in Appendix). Hence, the relatively stronger influence of heat release rate on the acoustic field has an important role in triggering the large oscillations of the acoustic field and consequently, TAI can sustain in the combustor.   

%$\textcolor{red}{DONE}$

\subsubsection{Mutual influence between acoustic pressure and vorticity fields}\label{subsubsec:acoustic-vorticity}

For a non-reacting cold flow, $p^{\prime}$ 
and flow dynamics characterized by vorticity field ($\omega$) and flow velocity oscillations ($u^{\prime}$) become correlated with each other when a flow passes through a duct~\cite{pedersen1997quantification}. Vorticity (curl of velocity field) is the intrinsic hydrodynamic characteristic of a flow system. In case of the reacting flows, for high temperature, baroclinic generation of vorticity increases~\cite{eftekharian2021analysis} due to an uneven torque acting on fluid components, such as when pressure and density gradients are not aligned. Therefore, vorticity generated in the liquid rocket engine combustor can be hydrodynamic as well as baroclinic. Harvazinski et al.~\cite{harvazinski2015coupling} showed that baroclinic vortex generation can be one of the crucial factors in self-excited instabilities in CVRC experiments. Also, the acoustic-flow interaction becomes more complex in such system since an additional effect comes as both the fields alter according to the variation in the feedback from $\dot{Q^{\prime}}$. Therefore, the mutual influence between $p^{\prime}$ and flow variables during their variations in the thermo-fluid systems differ from non-reactive cold flows. However, so far, we do not get a clear understanding of the mutual influence indicative of feedback interactions between $p^{\prime}$, $\omega$ and $u^{\prime}$ of the model liquid rocket engine combustors, including the present experimental setup. 

In the present study, we investigate the direction of the influence between the fluctuations in the acoustic and hydrodynamic flow fields by performing STE on $p^{\prime}$, $\omega$ and $u^{\prime}$ (results of $T^S_{u^{\prime}{\rightarrow}p^{\prime}}$ and $T^S_{p^{\prime}{\rightarrow}u^{\prime}}$ is discussed in Sec.~\ref{subsubsec:acoustic-velocity}) during $t=22.75$ to $22.80$ ms. 
%In the current section, we discuss the mutual influence between $p^{\prime}$ and $\omega$ in Fig.~\ref{fig_STE_simu_p_omega_acoustic rising}. Later, the mutual influence between $p^{\prime}$ and $u^{\prime}$ is discussed in Sec.~\ref{subsubsec:acoustic-velocity}. 
In the HAMSTER combustor, vortices are shed at the fuel collar from the oxidizer side~\cite{fuller2019dynamic}. Further, behind the fuel collar at the dump plane (backward facing step), we find the simultaneous rolling of positive (correspond to anti-clockwise vorticity) and negative (correspond to clockwise vorticity) vortices  at $t=22.75$ ms (see spatial distribution of $\omega$ in Fig.~\ref{fig_STE_simu_p_omega_acoustic rising} II (a)), signifying a development of low velocity regime (discussed in Sec.~\ref{subsubsec:acoustic-velocity}) over there. However, as the low velocity regime contracts after $t=22.80$ ms (discussed in Sec.~\ref{subsubsec:acoustic-velocity}), we see that the vortices eventually cease to revolve. 

Now, let us examine the influences of $p^{\prime}{\rightarrow}{\omega}$ and $\omega{\rightarrow}{p^{\prime}}$ (Figs.~\ref{fig_STE_simu_p_omega_acoustic rising} III and IV). We find that the influence of $p^{\prime}{\rightarrow}{\omega}$ mainly dominates in the downstream direction of the dump plane and the injector recess when $p^{\prime}$ reaches local maxima at $t=22.85$ ms. On the other hand, we notice that the influence of $\omega{\rightarrow}{p^{\prime}}$ spreads more rapidly near the dump plane than $\dot{Q}^{\prime}{\rightarrow}{p^{\prime}}$ (comparing Figs.~\ref{fig_STE_simu_pq_acoustic rising} IV and~\ref{fig_STE_simu_p_omega_acoustic rising} IV). 
%Thus, $\frac{d(\omega^{\prime}{\rightarrow}{p^{\prime}})}{dt} > \frac{d(\dot{Q}^{\prime}{\rightarrow}{p^{\prime}})}{dt}$ signifies that $\omega^{\prime}$ has greater impact on driving the acoustic field across the dump regime. 
In addition, the direction of influence $\omega{\rightarrow}p^{\prime}$ observed at the upstream direction of the dump plane through the injector recess and fuel collar implies that $\omega$ has a significant impact on $p^{\prime}$ that 
%causes
in turn influences both 1L mode and the bulk instability over there. 

\begin{sidewaysfigure*}%
\centering
\includegraphics[width=0.9\textwidth]{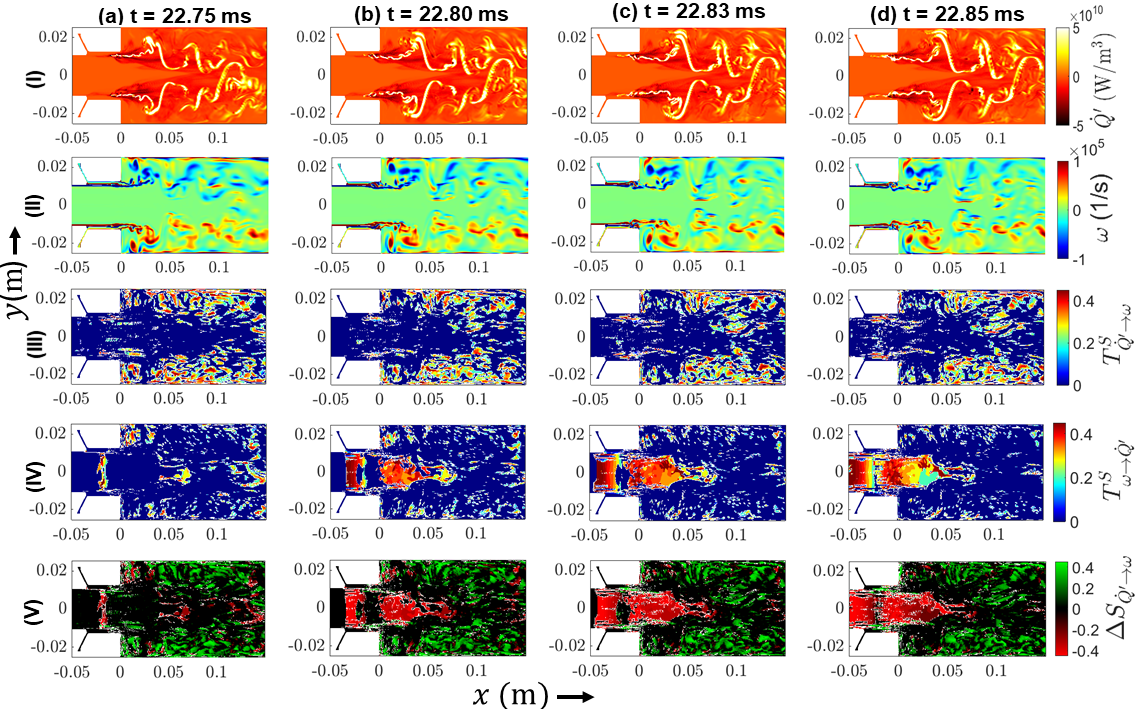}
\caption{The mutual influence between $\dot{Q}^{\prime}$ and $\omega$ during the amplification of the acoustic pressure near the dump plane is investigated. 
%Towards that, we consider four time steps that indicate the approaching of the system towards an acoustic rising: (a) $t=22.75$ ms, (b) $t=22.80$ ms, (c) $t=22.83$ ms and (d) $t=22.85$ ms. 
The distributions of $T^{S}_{{\dot{Q}^{\prime}}{\rightarrow}{\omega}}$ and $T^{S}_{{\omega}{\rightarrow}{\dot{Q}^{\prime}}}$ are shown at row-III and row-IV, respectively, during $t = 22.75$ to 22.85 ms. We find $\dot{Q}^{\prime}$ influencing zones near the wall of the combustor and also along the boundary of the flame where flame-vorticity interaction occurs (see $T^{S}_{{\dot{Q}^{\prime}}{\rightarrow}{\omega}}$). On the other side, we find the influencing zones of $\omega^{\prime}$ (see $T^{S}_{{\omega}{\rightarrow}{Q^{\prime}}}$) around the center of the combustor at the dump regime. From $\Delta S_{{\dot{Q}^{\prime}}{\rightarrow}{\omega}}$, we witness the relatively stronger influence of $\omega$ over ${\dot{Q}^{\prime}}$ at larger area near the entrance and the injector recess, while ${\dot{Q}^{\prime}}$ strongly influences $\omega$ near the wall of combustor. The dimension of combustor is considered in meter.}
\label{fig_STE_simu_Q_omega_acoustic rising}
\end{sidewaysfigure*}

From the distributions of $\Delta S_{{p^{\prime}}{\rightarrow}{\omega}}$ in Fig.~\ref{fig_STE_simu_p_omega_acoustic rising}, we notice a switching in the predominant direction of feedback interaction between $p^{\prime}$ and $\omega$ at the dump plane, and upstream and downstream of the combustor. At $t$ = 22.80 ms, we find that the switching of positive to negative values of $\Delta S_{{p^{\prime}}{\rightarrow}{\omega}}$ begins to occur near the fuel collar region, identifying the epicenter of the emergence of driving influences of vorticity field on the acoustic pressure field. 
After that, the rapid replacement of the positive values of $\Delta S_{{p^{\prime}}{\rightarrow}{\omega}}$ (red colored regions) by its negative values  (green colored areas) at the dump region and the upstream of the fuel collar, indicating the dominant direction of interaction from $\omega$ to $p^{\prime}$. On the other hand, the negative values of $\Delta S_{{p^{\prime}}{\rightarrow}{\omega}}$ are replaced by its positive values downstream of the combustor when ${p^{\prime}}$ reaches local maxima, signifying the relatively stronger influence of $p^{\prime}$ over $\omega$. Thus, the impact of the vortex shedding near the fuel collar on the amplification of $p^{\prime}$ is clearly captured through the directionality index. In literature, there is evidence of high-pressure oscillations sustained by vortex shedding in a dump combustor~\cite{knoop1997extension}. However, through the present study, understanding of a rapid switching in the driving direction of bidirectional coupling between $p^{\prime}$ and $\omega$ on spatiotemporal scales can help engineers in the control of TAI and redesign of the combustors.           

%a evolution of a cluster of negative $\Delta S_{{p^{\prime}}{\rightarrow}{\omega^{\prime}}}$ values throughout the dump region as well as at the upstream direction of the combustion inlet. On the other hand, the cells having positive values of $\Delta S_{{p^{\prime}}{\rightarrow}{\omega^{\prime}}}$ are seen to gradually shift towards the downstream direction of the dump regime, signifying that ${p^{\prime}}{\rightarrow}\omega^{\prime}$ driven dump regimes are also transformed into $\omega^{\prime}{\rightarrow}{p^{\prime}}$ driven regimes during the time period. The spreading of the driving regime of $\omega^{\prime}{\rightarrow}{p^{\prime}}$ towards the upstream as the acoustic amplifies provides information a priori about the rising pressure towards the upstream direction.  

%Additionally, we find the effective vorticity-acoustic interaction regimes that depend on the position as cwell as the dynamics of the vortices. Note that the vorticity is not zero at the upstream of the dump plane and the center of combustor, although small in magnitude ($|\omega|$) compared to the recirculation regimes (where, vortices shedding is seen) and the downstream of the dump plane. In short, a clear idea about the mutual influence between $p^{\prime}$ and $\omega^{\prime}$ can be obtained through the concept of symbolic transfer entropy, as shown in the figure.

\subsubsection{Mutual influence between heat release rate and vorticity fields}

\begin{figure*}%
\centering
\includegraphics[width=1.07\textwidth]{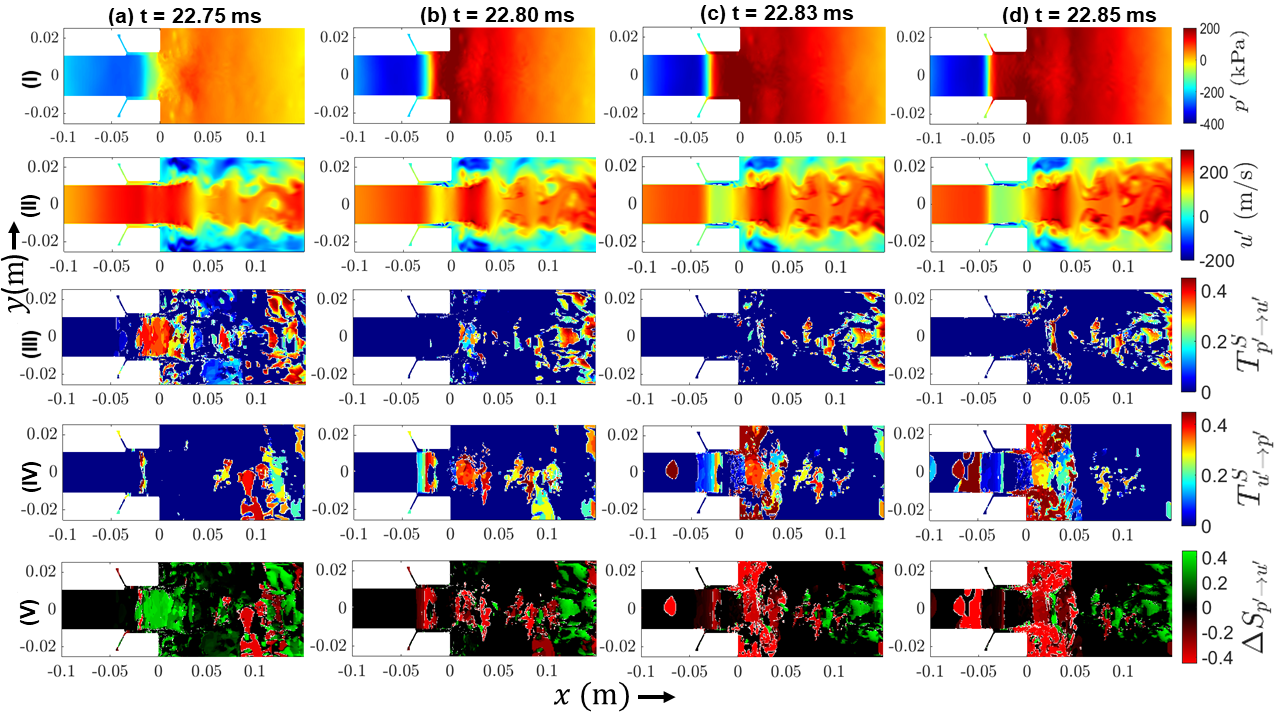}
\caption{The spatial distributions of $p^{\prime}$,  $u^{\prime}$, influence of $p^{\prime}{\rightarrow}u^{\prime}$ ($T^{S}_{{p^{\prime}}{\rightarrow}{u^{\prime}}}$), influence of $u^{\prime}{\rightarrow}p^{\prime}$ ($T^{S}_{{u^{\prime}}{\rightarrow}{p^{\prime}}}$) and the directionality ($\Delta S_{{p^{\prime}}{\rightarrow}{u^{\prime}}}$) are presented in separate rows at different time stamps: (a) $t$ = 22.75 ms, (b) $t$ = 22.80 ms, (c) $t$ = 22.83 ms and (d) $t$ = 22.85 ms. We notice  a strong influence of ${p^{\prime}}{\rightarrow}{u^{\prime}}$ ($T^{S}_{{p^{\prime}}{\rightarrow}{u^{\prime}}}$) near dump region which gradually dies down as the low velocity region diminishes near the wall. on the other hand, we find a gradual increase in the influence of ${u^{\prime}}{\rightarrow}{p^{\prime}}$ ($T^{S}_{{p^{\prime}}{\rightarrow}{u^{\prime}}}$) in the combustor. From $\Delta S_{{p^{\prime}}{\rightarrow}{u^{\prime}}}$ (corresponding to the fifth row), we identify the switching in the direction of predominant interaction between two fields. The combustor's dimension is shown in meter.}
\label{fig_STE_simu_p_uvelocity_acoustic rising}
\end{figure*}

Following the discussions about the mutual influences between $p^{\prime}$ and $\dot{Q}^{\prime}$ (Sec.~\ref{subsubsec:acoustic-heat}), and $p^{\prime}$ and $\omega$ (Sec.~\ref{subsubsec:acoustic-vorticity}), we try to find the direction of mutual influence (feedback interaction) between $\dot{Q}^{\prime}$ and $\omega$ in the current section. We present $T^{S}_{{\dot{Q}^{\prime}}{\rightarrow}{\omega}}$ and $T^{S}_{{\omega}{\rightarrow}{\dot{Q}^{\prime}}}$ in the third and fourth rows respectively in Fig.~\ref{fig_STE_simu_Q_omega_acoustic rising} to identify the zones where the direction of influence is seen along ${\dot{Q}^{\prime}}{\rightarrow}{\omega}$ and ${\omega}{\rightarrow}{\dot{Q}^{\prime}}$. We find the direction of influence along ${\dot{Q}^{\prime}}{\rightarrow}{\omega}$ near the wall and at the dump region of the combustion chamber. Consequently, we notice a significant variation in the distribution of $\omega$ at those areas. The previous study~\cite{harvazinski2020large} also showed that the increase in 
%heat release rate
${\dot{Q}^{\prime}}$ coincides with the timing of the impingement of the vortices at the wall. On the other hand, we can see the influence between ${\omega}$ and ${\dot{Q}^{\prime}}$ grows along ${\omega}{\rightarrow}{\dot{Q}^{\prime}}$ direction towards the injector recess (where fuel and oxidizes mix prior to the inlet of the combustion chamber)  %entrance of combustion chamber 
that coincides with the amplification in 
%heat release rate
$\dot{Q}^{\prime}$ (see 
%the distribution of $\dot{Q}$
in Fig.~\ref{fig_STE_simu_Q_omega_acoustic rising}) at the the injector recess. The vortex shedding from the fuel collar couples with the accelerated flow of hot gases at the injector recess, establishing a thin region of $\dot{Q}^{\prime}$ 
before near-complete combustion. 

Finally, we demarcate the regions of dominant direction of interaction in  ${\omega}{\rightarrow}{\dot{Q}^{\prime}}$ (red colored region) and ${\dot{Q}^{\prime}}{\rightarrow}{\omega}$ (green colored region) from $\Delta S_{{\dot{Q}^{\prime}}{\rightarrow}{\omega^{\prime}}}$ (Fig.~\ref{fig_STE_simu_Q_omega_acoustic rising} V). Notably, comparing the distribution of $\Delta S_{{\dot{Q}^{\prime}}{\rightarrow}{\omega}}$ with the distribution of $\Delta S_{{\dot{Q}^{\prime}}{\rightarrow}{p^{\prime}}}$ (Fig.~\ref{fig_STE_simu_pq_acoustic rising}), we can infer that the mechanism of TAI near the entrance to the combustion chamber is primarily driven by $\omega$ than $\dot{Q}^{\prime}$, whereas, $\dot{Q}^{\prime}$ strongly drives TAI than $\omega$ at the expanded area (i.e., backward step or dump plane) near the entrance. In addition, we understand that $\dot{Q}^{\prime}$ has stronger influence on $\omega^{\prime}$ near the wall than $p^{\prime}$. Understanding of the predominant influences of $\dot{p}^{\prime}$, $\dot{Q}^{\prime}$ and $\omega$ on spatiotemporal scales thus can help us in exploring the genesis of the combustion oscillations during TAI in the liquid rocket engine combustor. In addition, a comparative study in the relative influence of such variables, which is discussed later, can be helpful in taking preventive strategy for mitigating these oscillations.

%Now, we want to see the mutual influence between $Q^{\prime}$ and $\omega^{\prime}$ for the acoustic rising near dump plane during thermoacoustic oscillations. We find the combustion influencing zones near the wall and also along the boundary of the flame where flame-vortices interaction ocuurs (see the combustion influencing zones through $T^{S}_{{Q^{\prime}}{\rightarrow}{\omega^{\prime}}}$). On the other side, we find a strong vortex influencing zone (see $T^{S}_{{\omega^{\prime}}{\rightarrow}{Q^{\prime}}}$) around the center of the combustor at the dump regime. From $\Delta S_{{Q^{\prime}}{\rightarrow}{\omega^{\prime}}}$, We clearly find that the vortex driven regimes apperaing near dump plane and the injector recess, while combustion driven regimes are constantly seen near wall. It is noteworthy that the combustion oscillations have more influence on the vortices near the wall than the acoustic oscillations. Additionally, vortices have more influence on the combustion fluctuations close to the inlet than the acoustic.

\subsubsection{Mutual influence between acoustic pressure and flow velocity fields}\label{subsubsec:acoustic-velocity}

As a part of understanding the mutual influence between $p^{\prime}$ and hydrodynamic oscillations in the HAMSTER combustor, we perform STE analysis using spatiotemporal data of $p^{\prime}$ and $u^{\prime}$ %(fluctuations in flow velocity) 
in Fig~\ref{fig_STE_simu_p_uvelocity_acoustic rising}.   
From the spatiotemporal distributions of 
%flow velocity
$u^{\prime}$ (Fig.~\ref{fig_STE_simu_p_uvelocity_acoustic rising} II), we notice that the regions of high amplitude negative $u^{\prime}$ 
%velocity 
(reverse flow) regions, indicating the formation of the recirculation zones, are located close to the wall region of the combustor. 
However, when $p^{\prime}$ amplifies close to the dump plane, those negative velocity patches (blue-colored) that arise during the acoustic damping (when $p^{\prime}$ reaches local minima) begin to vanish. The positive $u^{\prime}$
%velocity 
(flow in the forward direction) regions (yellow and red-colored) progressively takes over the wall section during this period.  

\begin{sidewaysfigure*}%
\centering
\includegraphics[width=0.95\textwidth]{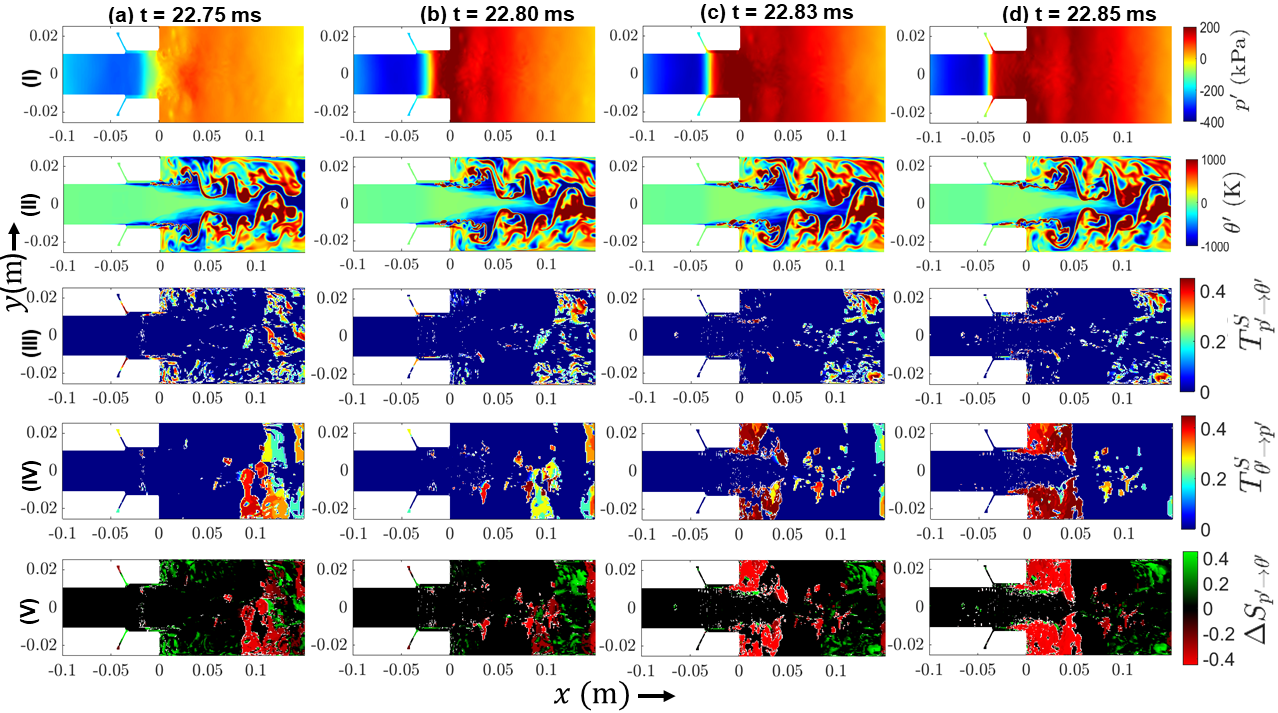}
\caption{The spatial distributions of $p^{\prime}$, $\theta^{\prime}$, influence of $p^{\prime}{\rightarrow}\theta^{\prime}$ ($T^{S}_{{p^{\prime}}{\rightarrow}{\theta^{\prime}}}$), influence of $\theta^{\prime}{\rightarrow}p^{\prime}$ ($T^{S}_{{\theta^{\prime}}{\rightarrow}{p^{\prime}}}$) and the directionality ($\Delta S_{{p^{\prime}}{\rightarrow}{\theta^{\prime}}}$) are shown in separate rows at different time stamps: (a) $t$ = 22.75 ms, (b) $t$ = 22.80 ms, (c) $t$ = 22.83 ms and (d) $t$ = 22.85 ms. We find high $T^{S}_{{p^{\prime}}{\rightarrow}{\theta^{\prime}}}$ consistently at the downstream of the combustor. On the other hand, prior to the local maxima of $p^{\prime}$, we observe a similar rise in $T^{S}_{{\theta^{\prime}}{\rightarrow}{u^{\prime}}}$ compared to $T^{S}_{{Q^{\prime}}{\rightarrow}{p^{\prime}}}$ at the dump region of the combustor. The dimension of the combustion is shown in meter.}
\label{fig_STE_simu_p_temp_acoustic rising}
\end{sidewaysfigure*}

In Figs.~\ref{fig_STE_simu_p_uvelocity_acoustic rising} III-IV, we try to identify the direction of influences between ${p^{\prime}}$ and ${u^{\prime}}$ on both spatial and temporal scales during the emergence of positive ${u^{\prime}}$ regions from the centerline to the wall ($t$ = 22.75 to 22.85 ms). 
We discover the influence of ${p^{\prime}}{\rightarrow}{u^{\prime}}$ (see $T^{S}_{{p^{\prime}}{\rightarrow}{u^{\prime}}}$ in Fig.~\ref{fig_STE_simu_p_uvelocity_acoustic rising} III) at the inlet of and at the dump region at $t=22.75$ ms and gradually declines after that, as the reverse flow events almost diminishes closer to the wall. On the other hand, we find that $T^{S}_{{p^{\prime}}{\rightarrow}{u^{\prime}}}$ persists with a little reduction in amplitude at the downstream of the dump plane where positive ${u^{\prime}}$ is continuously observed during the period.
This retention of $T^{S}_{{p^{\prime}}{\rightarrow}{u^{\prime}}}$ contributes to the expansion of the positive ${u^{\prime}}$ regions. In contrast, we see a gradual increase in the influence of ${u^{\prime}}{\rightarrow}{p^{\prime}}$ (i.e., $T^{S}_{{u^{\prime}}{\rightarrow}{p^{\prime}}}$) in Fig.~\ref{fig_STE_simu_p_uvelocity_acoustic rising} IV. 

Further, from $\Delta S_{{p^{\prime}}{\rightarrow}{u^{\prime}}}$ (Fig.~\ref{fig_STE_simu_p_uvelocity_acoustic rising} V[(a)-(b)]), we clearly detect that the direction of dominant interaction between $p^{\prime}$  and $u^{\prime}$  switches from ${p^{\prime}}{\rightarrow}{u^{\prime}}$ to ${u^{\prime}}{\rightarrow}{p^{\prime}}$ near the fuel collar at $t$ = 22.80 ms. After that, the driving influence of $u^{\prime}$ on $p^{\prime}$ can be progressively seen stronger near the dump region.  The predominance direction of the interaction in ${u^{\prime}}{\rightarrow}{p^{\prime}}$ indicates that the formation of a positive velocity region is one of the driving phenomena for the amplification in $p^{\prime}$. Thus, in general, observing the switching of the driving direction in $\omega{\rightarrow}{p^{\prime}}$ (Sec.~\ref{subsubsec:acoustic-vorticity}) and ${u^{\prime}}{\rightarrow}{p^{\prime}}$, we can mention that the switching of the predominance direction of hydrodynamic-acoustic interaction in hydrodynamic${\rightarrow}$ acoustic near the dump region is one of the key factors in the rise of $p^{\prime}$.

\subsubsection{Mutual influence between acoustic pressure and temperature of combustion fields }

%To understand the mutual influence between the fluctuation in temperature ($\theta^{\prime}$) due to the burning of propellants during combustion and $p^{^\prime}$, we first present the spatio-temporal distributions of these two variables at first two rows for easier understanding. We notice a lower temperature zone at the wall close to the dump region where the heat exchange between unburnt and burnt propellants occurs. However, we find that the expansion of high temperature zones at the downstream of dump plane as consequence of rapid combustion of the propellants when acoustic approaches maxima. From $T^{S}_{{p^{\prime}}{\rightarrow}{\theta^{\prime}}}$ (see spatio-temporal distributions at third row), we explore that the fluctuation in $p$ has influence on $\theta^{\prime}$ at the downstream since islands of $T^{S}_{{p^{\prime}}{\rightarrow}{\theta^{\prime}}}$ can be constantly seen at there. On the other hand, we unveil that $T^{S}_{{\theta^{\prime}}{\rightarrow}{u^{\prime}}}$ increases near the dump plane similar to $T^{S}_{{Q^{\prime}}{\rightarrow}{p^{\prime}}}$ as the burning temperature can be directly influenced by the combustion kintetics.
Like other variables discussed in the previous sections, the knowledge of the influence of the burning rate of propellants on the acoustic pressure waves is also required to suppress combustion instabilities in the rocket engine combustors~\cite{kathiravan2021acoustic}. 
The temperature of the combustion field is primarily linked with the density of unburnt reactants and products, the perturbations in the burning area due to equivalence ratio disturbances~\cite{lieuwen2021unsteady}, and proportional to the burning rate of the propellants~\cite{lengelle2002combustion}. 

To understand the mutual interaction between $p^{\prime}$ and $\theta^{\prime}$ (fluctuations in temperature of the combustion field) during the approach towards a local maxima of $p^{\prime}$ during TAI, we need to identify the direction of influences between these two variables on both spatial and temporal scales.   
Towards that, we first present the spatio-temporal distributions of these two variables (Fig.~\ref{fig_STE_simu_p_temp_acoustic rising} I-II) at the first two rows followed by the STE analysis (Fig.~\ref{fig_STE_simu_p_temp_acoustic rising} III-V). We notice the lower temperature zones at the wall close to the dump region where the heat exchange between unburnt propellants and products gradually occurs. The heat transfer between the unburnt and burnt mixtures takes place when high temperature burnt gases move to the recirculation zone (i.e., negative velocity region) and consequently, fuel is rapidly consumed there. On the other hand, the high temperature zones are confined in the shear layer near the dump region. 
%However, the higher temperature zones are expanded to the wall region at the downstream of the dump plane. 
Further, we observe that a slow expansion of high temperature zones at the downstream of the dump plane promotes the rapid combustion of the propellants when $p^{\prime}$ approaches a local maxima.    

From $T^{S}_{{p^{\prime}}{\rightarrow}{\theta^{\prime}}}$ (see Fig.~\ref{fig_STE_simu_p_temp_acoustic rising} III), we explore that $p^{\prime}$ influences $\theta^{\prime}$ downstream of the combustor since the islands of $T^{S}_{{p^{\prime}}{\rightarrow}{\theta^{\prime}}}$ can be constantly seen there. On the other hand, we unveil that $T^{S}_{{\theta^{\prime}}{\rightarrow}{p^{\prime}}}$ increases near the dump plane similar to $T^{S}_{{\dot{Q}^{\prime}}{\rightarrow}{p^{\prime}}}$ 
%since the chemical kinetics of intermediate species, such as CH$^{*}$, OH$^{*}$, etc., can directly influence the combustion temperature~\cite{turns1996introduction}
when the high temperature zone expands to the dump plane and enhances the heat transfer with the unburnt mixture. 
Observing the negative and positive values of $\Delta S_{{p^{\prime}}{\rightarrow}{\theta^{\prime}}}$ (Fig.~\ref{fig_STE_simu_p_temp_acoustic rising} V), we can also demarcate the areas where $p^{\prime}-\theta^{\prime}$ feedback interaction is  driven in $\theta^{\prime}{\rightarrow}p^{\prime}$ and $p^{\prime}{\rightarrow}\theta^{\prime}$ directions. We observe that the switching of $p^{\prime}{\rightarrow}\theta^{\prime}$ to $\theta^{\prime}{\rightarrow}p^{\prime}$ occurs at dump region, while the dominant direction of the interaction is seen in $p^{\prime}{\rightarrow}\theta^{\prime}$ downstream of the combustor near the local maxima of $p^{\prime}$.

Thus, we find a clear perception of the dominant direction of influences between %$p^{\prime}$, $\dot{Q}^{\prime}$, $\omega^{\prime}$, $u^{\prime}$ and $\theta^{\prime}$ 
the acoustic, flame and flow variables on spatiotemporal scales during an amplification of acoustic pressure oscillations in the HAMSTER combustor, while knowledge based on temporal data may be insufficient for understanding the sustenance mechanism of TAI. In the study using spatiotemporal variations of the variables employing STE, identifying the dominating feedback direction makes it simpler to find the relatively stronger influencing variable in a bidirectional interaction. A recent industrial case study~\cite{bauer2004specifying} using the concept of directionality ascertained the impact of a temperature controller on the downstream temperature readings and identified a regularity in the flow of influences for consecutive industrial processes. This work illustrated how the directionality measure can provide a precise representation of the dependencies within a process.    

In the present study, even though the qualitative analysis (Sec.~\ref{subsec:STE on spatiotemporal}) based on STE yields some conclusive results about the dominance between acoustic and other variables, we still need to find a robust way to identify the most influencing variables among all the variables for each time instant. 
%Therefore, we need to compare the influences of all variables to quantitatively understand the switching in the driving mechanism during a limit cycle of TAI.} 
In the following section (Sec.~\ref{subsec:STE_quantification}), we perform a quantification analysis based on STE to understand the switching in the driving mechanism at each time instant and determine the order of variables according to their impact of influences during a limit cycle of TAI.  

\subsection{Quantification of the influences between the variables during TAI}\label{subsec:STE_quantification}

\begin{figure*}%
\centering
\includegraphics[width=1.0\textwidth]{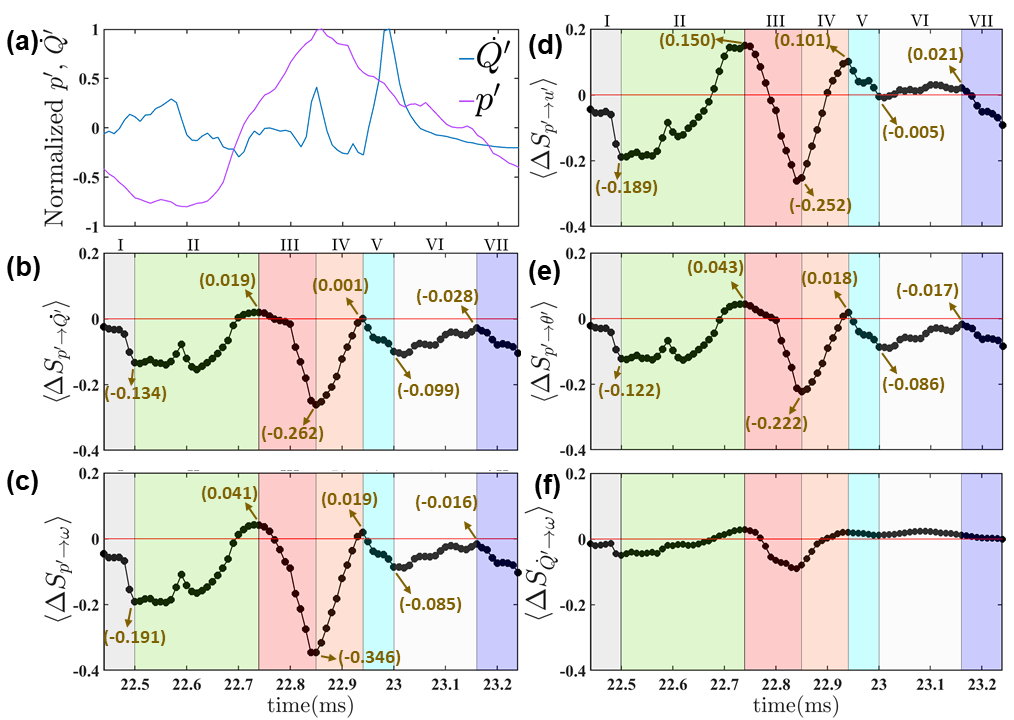}
\caption{A quantitative estimation of the degree of mutual influences among the fluctuations in acoustic field, hydrodynamic flow, vorticity, heat release rate and thermodynamics property such as temperature is provided in (b)-(f). 
%Identifying the significant spatiotemporal variations in mutual information around the dump plane during limit cycle oscillation (LCO), 
From the quantification, we are interested in finding an order of the primary variables according to their impact of influence. In this analysis, we choose a location close to the dump plane considering $0 \le x \le 0.05$ m, $-0.0125$ m $\le y \le$ 0.0125 m, and $z=0$ to calculate the spatially averaged directionalities ($\langle\Delta S\rangle$) in (b)-(f). Based on the variation of $\langle\Delta S\rangle$, we find seven distinct stages. In each stage, we find out the dominant direction of feedback interactions and the driving variable.}
\label{fig_STE_quantification}
\end{figure*}

Following the spatiotemporal variations of the directionality between the variables, here, we attempt to provide a quantitative estimation of the mutual influences between them during a cycle of TAI. We can understand that these variables influence each other autonomously during a combustion process. Hence, a linear correlation is insufficient to quantify the mutual information between the variables. The proposed STE enables one to consider the information flowing from one variable to another variable, while multiple variables ($\ge 2$) take part in the process. Thus, the approach provides an easier way to distinguish the mutual influences imposed by any variable on another variable in a nonlinear system. For understanding the temporal variation of the acoustic field and the heat release rate, we present $p^{\prime}$ and $\dot{Q}^{\prime}$ in Fig.~\ref{fig_STE_quantification}. %corresponding to $x$ = 0.0584 m and $0.010$ m $\le y \le$ 0.0125 m (location of the acoustic sensor of the experimental setup)
During TAI, we observe that $\dot{Q^{\prime}}$ intermittently exhibits the local maxima and minima that correspond to the attachment and detachment of the flame edge to the injector rim.  
%In such scenarios, estimation of phase between $p$ and $Q$ may not be helpful in understanding the mechanism between acoustic and combustion. 
%Finding the significant spatiotemporal variations in the mutual information near the dump plane during limit cycle oscillation (LCO), we are interested identify the order of the primary variables according to their impact of influence. Towards that, 

Further, we select an area considering $0 \le x \le 0.05$ m, $-0.0125$ m $\le y \le 0.0125$ m, and $z=0$ to calculate the spatially averaged directionalities ($\langle\Delta S\rangle$) in Figs.~\ref{fig_STE_quantification} (b)-(f). The region of interest (ROI) is selected such that it coincides with the switching in the direction of the feedback interactions between the variables in the combustion chamber near the dump region.
By calculating $\langle\Delta S\rangle$ between two variables and comparing all $\langle\Delta S\rangle$ with each other, we can identify the driving or most influencing variables during one cycle of oscillation. For easier understanding of the impact of relative influences between two variables, we use separate plots, (b)-(f) in the figure. 
%Readers can also look at the supplementary figure where these all graphs are shown in single plot to find a quantitative comparison among the influences. 

Based on the trend of $\langle\Delta S\rangle$ in Figs.~\ref{fig_STE_quantification} (b)-(e), we identify seven distinct stages of feedback interactions between the variables, each of them exhibits different mechanism. When $p^{\prime}$ drops near dump region during $t=22.44-22.5$ ms, we find that $\langle\Delta S\rangle$ in all cases, shown in the Figs.~\ref{fig_STE_quantification} (b)-(e), has the same trend below zero, although those are different in the amplitude. The negative values of $\langle\Delta S\rangle$ for those cases suggest that the relative influences of $\dot{Q}^{\prime}$, $\omega$, $u^{\prime}$ and $\theta^{\prime}$ on $p^{\prime}$ is stronger during this period. We consider this period as stage-I. At this stage, we find that $\langle\Delta S_{{p^{\prime}}{\rightarrow}{\omega}}\rangle \simeq \langle\Delta S_{{p^{\prime}}{\rightarrow}{u^{\prime}}}\rangle < \langle\Delta S_{{p^{\prime}}{\rightarrow}{\dot{Q}^{\prime}}}\rangle < \langle\Delta S_{{p^{\prime}}{\rightarrow}{\theta^{\prime}}}\rangle$ (Figs.~\ref{fig_STE_quantification} (b)-(e)). $\langle\Delta S_{{p^{\prime}}{\rightarrow}{\omega}}\rangle \simeq \langle\Delta S_{{p^{\prime}}{\rightarrow}{u^{\prime}}}\rangle$ signifies that the impact of influence of $\omega$ and $u^{\prime}$ on $p^{\prime}$ is nearly the same. In addition,  from $\langle\Delta S_{{p^{\prime}}{\rightarrow}{\omega}}\rangle < \langle\Delta S_{{p^{\prime}}{\rightarrow}{\dot{Q}^{\prime}}}\rangle$ (Figs.~\ref{fig_STE_quantification} (b)-(c)) and $\langle\Delta S_{{\dot{Q}^{\prime}}{\rightarrow}{\omega^{\prime}}}\rangle$ < 0 (Fig.~\ref{fig_STE_quantification} (f)), we understand that impact of influence of $\omega$ is higher than that of $\dot{Q}^{\prime}$. During this period, a previous study~\cite{harvazinski2020large} found that the overall heat release rate in the shear layer becomes thin and low overall. On the other hand, the shed vortex is headed straight towards the combustor wall at this time since the shear layer expands outward. We can make an order of the variables based on the impact of the relative influences: $\omega \simeq u^{\prime} > \dot{Q}^{\prime}>\theta^{\prime}>p^{\prime}$. Note that the order is found based on the influences of the acoustic pressure on other variables and vice versa.  

During $t = 22.5-22.74$ ms, we find that $\langle\Delta S_{{p^{\prime}}{\rightarrow}{u^{\prime}}}\rangle$ gradually increases and obtains the highest positive value followed by $\langle\Delta S_{{p^{\prime}}{\rightarrow}{\theta^{\prime}}}\rangle$ and $\langle\Delta S_{{p^{\prime}}{\rightarrow}{\dot{Q}^{\prime}}}\rangle$. We consider this time period as stage-II. We also notice that $\langle\Delta S_{{p^{\prime}}{\rightarrow}{\theta^{\prime}}}\rangle \simeq \langle\Delta S_{{p^{\prime}}{\rightarrow}{\omega}}\rangle$ at the end of stage-II. From this investigation, we unravel that $p^{\prime}$ is strongly influencing the other variables and has highest (lowest) driving influences on $u^{\prime}$ ($\dot{Q}^{\prime}$). Previous study~\cite{harvazinski2020large} found an overall thickening of the flame region in the shear layer for which the relative influence of $\dot{Q}^{\prime}$ on $\omega$ is found to be higher during this period. The dominant influencing order of the variables is arranged during stage-II as follows: $p^{\prime} >\dot{Q}^{\prime}>\omega\simeq\theta^{\prime}>u^{\prime}$. 

During $t=22.74-22.85$ ms, we find a decreasing trend of $\langle\Delta S\rangle$ that exhibits large negative values after crossing zero. We regard this period as stage-III. During this stage, we find the largest negative value of $\langle\Delta S_{{p^{\prime}}{\rightarrow}{\omega}}\rangle$ followed by $\langle\Delta S_{{p^{\prime}}{\rightarrow}{\dot{Q}^{\prime}}}\rangle$, $\langle\Delta S_{{p^{\prime}}{\rightarrow}{u^{\prime}}}\rangle$ and $\langle\Delta S_{{p^{\prime}}{\rightarrow}{\theta^{\prime}}}\rangle$. From this estimation, we unravel that the most influencing variable is $\omega$, while the relative influence of $p^{\prime}$ on the other variables is lower. From $\langle\Delta S_{{p^{\prime}}{\rightarrow}{\omega}}\rangle$ < $\langle\Delta S_{{p^{\prime}}{\rightarrow}{\dot{Q}^{\prime}}}\rangle$, we understand that the vortex-shedding events between the wall and shear layer mainly dominate the dynamics of the combustor, although the heat release rate becomes high. During stage-III, we make the following order of variables according to their influencing ability: $\omega > \dot{Q}^{\prime} > u^{\prime} > \theta^{\prime} > p^{\prime}$.

%\textcolor{red}{DONE}

We find that the trend of $\langle\Delta S\rangle$ in Figs.~\ref{fig_STE_quantification} (b)-(e) increases rapidly than stage-II during $t = 22.85-22.94$ ms and exhibits positive value at the end of this period. Finding such alteration in $\langle\Delta S\rangle$, we consider this time period as stage-IV. During this period, we also observe that $\langle\Delta S_{{p^{\prime}}{\rightarrow}{\dot{Q}^{\prime}}}\rangle \simeq 0$. Hence, the mutual influence between ${p^{\prime}}$ and ${\dot{Q}^{\prime}}$ becomes almost equal, while the relative influence of ${p^{\prime}}$ on the other variables is stronger. Additionally, we notice that $\langle\Delta S_{{p^{\prime}}{\rightarrow}{\omega}}\rangle$ becomes equal with $\langle\Delta S_{{p^{\prime}}{\rightarrow}{\theta^{\prime}}}\rangle$, signifying that the driving influences of both $\omega$ and $\theta^{\prime}$ on $p^{\prime}$  are equal. Further, we find that $\langle\Delta S_{{\dot{Q}^{\prime}}{\rightarrow}{\omega}}\rangle$ > 0 at the end of stage-IV, indicating a relatively stronger influence of ${\dot{Q}^{\prime}}$ on ${\omega}$. Based on those observations, we find the order of the variables according to their impact of relative influences at the end of stage-IV: $p^{\prime} \simeq \dot{Q}^{\prime}>\theta^{\prime}\simeq \omega>u^{\prime}$. 

During $t=22.94-23$ ms, we notice that $\langle\Delta S\rangle$ becomes negative in all cases (see Figs.~\ref{fig_STE_quantification} (b)-(e)). During this period, $\langle\Delta S_{{p^{\prime}}{\rightarrow}{\dot{Q}^{\prime}}}\rangle$ obtains the largest negative value, signifying that $\dot{Q}^{\prime}$ strongly influences the dynamics of the combustor. We also find a local maxima in $\dot{Q}^{\prime}$ at $t = 23$ ms (see in Fig.~\ref{fig_STE_quantification} a).  We consider the time period during 22.94-23 ms as stage-V. At the end of this stage, $\langle\Delta S_{{p^{\prime}}{\rightarrow}{\theta^{\prime}}}\rangle$ is found to be equal with $\langle\Delta S_{{p^{\prime}}{\rightarrow}{\omega}}\rangle$, indicating that the driving influences of thermal oscillations from combustion zone and vortex perturbations are the same during this stage. Besides, we observe that the driving influence of $u^{\prime}$ on $p^{\prime}$ is relatively lower than other variables as $\langle\Delta S_{{p^{\prime}}{\rightarrow}{u^{\prime}}}\rangle$ > ($\langle\Delta S_{{p^{\prime}}{\rightarrow}{\dot{Q}^{\prime}}}\rangle$, $\langle\Delta S_{{p^{\prime}}{\rightarrow}{\omega}}\rangle$, $\langle\Delta S_{{p^{\prime}}{\rightarrow}{\theta^{\prime}}}\rangle$). We note a order during this phase as follows: $\dot{Q}^{\prime} > \theta^{\prime}\simeq \omega>u^{\prime}>p^{\prime}$.  

After stage-V, we find an almost constant behavior of $\langle\Delta S\rangle$ till $t= 23.16$ ms. We regard this time period as stage-VI. During this stage, we find that $\langle\Delta S_{{p^{\prime}}{\rightarrow}{u^{\prime}}}\rangle$ > 0, signifying that the relative influence of ${p^{\prime}}$ on ${u^{\prime}}$ is stronger. On the other hand, ($\langle\Delta S_{{p^{\prime}}{\rightarrow}{\dot{Q}^{\prime}}}\rangle$, $\langle\Delta S_{{p^{\prime}}{\rightarrow}{\theta^{\prime}}}\rangle$, $\langle\Delta S_{{p^{\prime}}{\rightarrow}{\omega}}\rangle$) < 0, signifying that the relative influences of ${\dot{Q}}^{\prime}$, $\omega$ and $\theta^{\prime}$ on ${p^{\prime}}$ is stronger. Further, we understand that ${\dot{Q}^{\prime}}$ strongly influences ${\omega}$ from $\langle\Delta S_{{\dot{Q}^{\prime}}{\rightarrow}{\omega}}\rangle > 0$ (in Fig.~\ref{fig_STE_quantification} f)  and $\langle\Delta S_{{p^{\prime}}{\rightarrow}{\omega}}\rangle > \langle\Delta S_{{p^{\prime}}{\rightarrow}{\dot{Q}^{\prime}}}\rangle$  (in Figs.~\ref{fig_STE_quantification} (b)-(c)). Besides, we notice that $\langle\Delta S_{{p^{\prime}}{\rightarrow}{\omega}}\rangle \simeq \langle\Delta S_{{p^{\prime}}{\rightarrow}{\theta^{\prime}}}\rangle < 0$, signifying that the driving influences from both $\omega$ and $\theta^{\prime}$ becomes same at the end of stage-VI. Thus, we can find the driving order of the variables during stage-VI as follows:   $\dot{Q}^{\prime}> \theta^{\prime}\simeq \omega>p^{\prime}>u^{\prime}$.

Prior to the end of one cycle during TAI, we find a change in trend of $\langle\Delta S\rangle$ during $t=23.16-23.24$ ms. We regard this period as stage-VII. Interestingly, we find the difference between values of $\langle\Delta S\rangle$ of the variables (Figs.~\ref{fig_STE_quantification} (b)-(e)) is small towards $t=23.24$ ms. This signifies that the driving influences of $\dot{Q}^{\prime}$, $\omega$, $u^{\prime}$, $\theta^{\prime}$ on $p^{\prime}$ is more or less same.

In a nutshell, we unravel the variation in the predominant direction of the feedback interactions among the primary variables of a liquid rocket engine combustor during one cycle of thermoacoustic oscillation based on the variation of the directionality index. 
Further, by quantifying the different levels of feedback interactions between the %combustion, acoustic field, and hydrodynamic variables, we obtain a clear perception of the dominant direction of the feedback interactions between them in a highly complex reacting flow system.   
%Moreover and most importantly, 
variables, we find the driving order of the variables according to the impact of their influence or feedback and witness the switching of the driving order, causing the change in the dynamics of the combustor. From the spatiotemporal behavior and the quantification analysis, we unveil that the vorticity emerging from the fuel collar is the most influencing variable for the acoustic pressure fluctuation approaching a local minima (stage-I) or maxima (stage-III). The identification of the fuel collar as the epicenter of TAI can be helpful in mitigation of the instability by adopting appropriate strategy such as the suppression of vortex shedding using flow modification methods~\cite{krishnappa2022aerodynamic,rashidi2016vortex}. Also, the understanding the behavior of stage-II and stage-III (also see $\langle\Delta S\rangle$ between $p^{\prime}$ and $\dot{Q}^{\prime}$ for considering different cycle of oscillation during TAI in Appendix C) is crucial since that period is close to the local maxima of $p^{\prime}$. Thus, the symbolic transfer entropy is useful approach to examine the relative influence of the bidirectional coupling between the variables in understanding the physics underlying the sustenance of thermoacoustic instability. Nevertheless, the importance of symbolic transfer entropy is not limited to the HAMSTER combustor and can be useful to understand the feedback interaction between the system variables at different dynamical states for the turbulent combustors.      
%General inferences from the results

\section{Conclusions}\label{sec:conclusions}

%We use a dataset obtained from a high-fidelity large eddy simulation (LES) of a first longitudinal (1L) mode dominated liquid rocket engine combustor to understand the direction of the feedback interaction between combustion, acoustic field and hydrodynamic variables during the occurrence of thermoacoustic instability. Towards this, we use an approach that combines transfer entropy with symbolic dynamic filtering to analyze the influence of one variable on the other variables in a liquid rocket engine combustor called HAMSTER. Under this framework, we measure the influence between the variables in both directions and identify the dominant route of influence between them. Based on the experimental data of OH* chemiluminescence and acoustic pressure oscillations at the combustor wall close to the dump regime, we notice a continuous decline in the influence of OH* on acoustic field, leading to a considerably stronger acoustic influence on the reaction rate. After confirming that our observation based on temporal data agrees with the results reported in earlier literature, we move forward with a thorough analysis of the spatiotemporal behavior of the variables related to the combustion and hydrodynamic fields.

We utilize a dataset derived from a high-fidelity large eddy simulation (LES) of a first longitudinal (1L) mode dominated liquid rocket engine combustor to investigate the feedback interaction between combustion, acoustic field, and hydrodynamic variables during the occurrence of thermoacoustic instability. In this study, we employ an approach that integrates transfer entropy with symbolic dynamic filtering to analyze the influence of one variable on the others in a liquid rocket engine combustor known as HAMSTER. Within this framework, we measure the influence between the variables in both directions and identify the dominant route of influence between them. Using experimental data of OH* chemiluminescence and acoustic pressure oscillations at the combustor wall near the dump regime, we observe a continuous decrease in the influence of OH* on the acoustic field, leading to a notably stronger acoustic influence on the reaction rate. Upon confirming that our temporal data observations align with previous literature, we proceed with a comprehensive analysis of the spatiotemporal behavior of the variables related to the combustion and hydrodynamic fields.

%In the spatiotemporal study, we discuss the mutual influences and the directionality (i.e., net influence between the variables) between the most prominent variables: acoustic pressure, heat release rate, vorticity, flow velocity, and temperature when the fluctuation of acoustic pressure reaches the local maxima of a cycle during thermoacoustic oscillations. We observe that the influence of all variables on the acoustic field significantly increases during this period. Based on the spatiotemporal analysis using symbolic transfer entropy, we clearly identify the influencing islands of the variables at each time stamp. We discover that the heat release rate oscillations influence the acoustic pressure oscillations on either side of the shear layer at the dump plane. On the other hand, the influence of acoustic pressure oscillations on the heat release rate oscillations observed at the dump region gradually diminishes. Further, we detect that the influence of the fluctuations in vorticity on the acoustic field starts near the fuel collar when acoustic pressure approaches a local maxima. The influence of vorticity on the acoustic pressure oscillations is found throughout the dump region and extends toward the upstream region of the dump plane during this time.
In our spatiotemporal study, we examine the mutual influences and the directionality (i.e., net influence between the variables) among key factors: acoustic pressure, heat release rate, vorticity, flow velocity, and temperature, specifically when the fluctuation of acoustic pressure reaches the local maximum within a thermoacoustic oscillation cycle. We observe a significant increase in the influence of all variables on the acoustic field during this period. Through spatiotemporal analysis using symbolic transfer entropy, we are able to clearly identify the influencing "islands" of the variables at each time stamp. Our findings reveal that heat release rate oscillations exert influence on acoustic pressure oscillations on both sides of the shear layer at the dump plane, with the influence of acoustic pressure oscillations on heat release rate oscillations in the dump region gradually diminishing. Furthermore, we detect that the influence of vorticity fluctuations on the acoustic field commences near the fuel collar as the acoustic pressure approaches a local maximum. This influence extends throughout the dump region and towards the upstream region of the dump plane during this time.
%The relatively stronger influence of vorticity in the upstream direction contributes to the rise in acoustic pressure there. We also understand the influence between heat release rate oscillations and vorticity and demarcate the most influencing zones of those variables on the acoustic pressure oscillations. The influence of heat release rate oscillations is found to be predominant than that of vorticity on the acoustic field at the expanded area of the dump plane close to the fuel injector and the wall of the combustor. On the other hand, the influence of vorticity is found to be stronger than that of heat release rate oscillations in the injector recess and around the center of the combustor. The influence of velocity on the acoustic field increases near the dump region and the fuel collar when a positive velocity zone gradually replaces a negative velocity zone at the wall of the combustor. We also notice that the effect of temperature oscillations of combustion field on the acoustic field increases during the movement of the high-temperature zone toward the dump region. 
The relatively stronger influence of vorticity in the upstream direction contributes to the rise in acoustic pressure there. We also understand the influence between heat release rate oscillations and vorticity and demarcate the most influencing zones of those variables on the acoustic pressure oscillations. The influence of heat release rate oscillations is found to be predominant than that of vorticity on the acoustic field at the expanded area of the dump plane close to the fuel injector and the combustor wall. On the other hand, vorticity has a stronger influence than heat release rate oscillations in the injector recess and around the center of the combustor. The influence of velocity on the acoustic field increases near the dump region and the fuel collar, especially when a positive velocity zone gradually replaces a negative velocity zone at the combustor wall. Additionally, we observed that the effect of temperature oscillations in the combustion field on the acoustic field intensifies as the high-temperature zone moves toward the dump region.

%Further, from the estimation of the directionality index, we clearly notice an alteration in driving near the dump region when acoustic pressure rises for all cases.}

%\textcolor{red}{(You can even write points and give conclusions from temporal STE and spatiotemporal STE separately (making it clear).}

Further, we compute the spatially averaged directionalities between the acoustic pressure oscillations and the other variables to quantify and compare their level of influence. 
%The variation in the directionalities indicates that the driving mechanism changes after a particular time duration during one cycle of thermoacoustic oscillation due to the variation in the level of influence of the variables. 
We observe that the driving mechanism changes at specific points in the thermoacoustic oscillation cycle due to the varying influence of these variables. The analysis quantifies that the vorticity is the stronger influencing variable compared to others during an amplification (a reduction) of acoustic pressure to a local maxima (minima). The influence of velocity field is more or less equal to the vorticity toward the local minima. This way, the study explicitly 
%discusses
provides the insights into
the relative influences of all possible physical variables in sustaining thermoacoustic instability for the first time. The perception of the switching in the predominance of the direction of feedback interactions during local minima to the local maxima of acoustic pressure oscillations based on CFD simulation data will help the researchers during the design stage of the combustors. Further, the strategy can be adopted in the design to target the epicenter of the emergence of thermoacoustic instability in the rocket engine combustor. Finally, this symbolic transfer entropy-based analysis can be a suitable approach for exploring the driving mechanism of various reacting flow systems extending beyond the %1L mode rocket combustor
axial mode dynamics of the model rocket combustor.

\begin{acknowledgments}
R. I. S acknowledges the funding provided by the Office of Naval Research Global (Grant No. N62909-22-1-2011, contract monitor: Dr. Derrick Marcus Tepaske). S.D. would also like to acknowledge the Post Doctoral Researcher fellowship under the same grant supported by the Office of Naval Research Global at Indian Institute of Technology Madras. We thank 
%Mr. Michael Orth and Professor Timothee Pourpoint 
Dr. Tristan Fuller for performing the experimental campaign at Purdue University. Distribution Statement A.  Approved for public release: distribution is unlimited AFRL-2024-2875. R.I.S. thanks Dr. Venkateswaran Sankaran (AFRL) for his numerous productive discussions. S.D. thanks to Dr. Ankit Sahay and Dr. Ankan Banerjee for stimulating discussion on the analysis. R.I.S, S.D. and P.K. thank to Mr. Vivek J and Mr. Chavaly from kfourmetrics for helping in obtaining the simulation data in the required format.   
\end{acknowledgments}

\section*{Data Availability Statement}

The data that support the findings of this study are available from the corresponding author upon reasonable request.

\section*{Declarations: Conflict of interests}
The authors declare no competing interests.

%\section*{Author Contributions}
 
\section*{Appendix}
\appendix
%\section{Estimation of Mach number for binary cyclones}\label{sec:A10}
%\textcolor{blue}{For the binary cyclone systems, we calculate the maximum Mach number in the following way:
%\begin{equation}
%\label{eqn:6}
%M_{cyclone}=\underset{u,v,w}{\mathrm{max}}\frac{\sqrt{u^2+v^2+w^2}}{\sqrt{\gamma*R*T}}
%\end{equation}
%Here, $u$($v$) is the horizontal component of the wind velocity to the east (north), $w$ is the velocity of the wind in an upward or downward direction, $\gamma$ is the heat capacity ratio or adiabatic index, $R$ is the specific gas constant and  $T$ is the air temperature at 850 hPa. Using $u$, $v$, $w$ from ERA5 data and choosing $R$ = 287 Jkg$^{-1}$K$^{-1}$ and $\gamma$ = 1.4, we obtain $M_{cyclone}$ to be 0.15 for the Seroja-Odette and Noru-Kulap pairs.} 

%\section{Calculation steps for estimating the separation distance}\label{sec:A11}

%Following Eqs. (\ref{eqn:7}-\ref{eqn:9}) are used to calculate the separation distance between two nearby cyclones in the present work.

%\begin{equation}
%\label{eqn:7}
%B_1=\sin^2{\frac{\delta \phi}{2}}+\cos{\phi_{1}}*\cos{\phi_{2}}*\sin^2{\frac{\delta \theta}{2}}
%\end{equation}

%Here, $\phi_1$ and $\phi_2$ are the latitudes of two cyclones at a particular time instance. We calculate the difference in latitude, $\delta \phi = \phi_1 - \phi_2$. Similarly, $\delta \theta$ is the difference in longitude of two corresponding cyclones.

%\begin{equation}
%\label{eqn:8}
%B_2=2\tan^{-1}(\sqrt{B1}, \sqrt{1-B1}) 
%\end{equation}

%\nocite{*}
\subsection*{A. Power spectrum at 180.34 mm downstream of the dump plane}\label{Appendix:A}

\begin{figure*}[ht!]%
\centering
\includegraphics[width=1.02\textwidth]{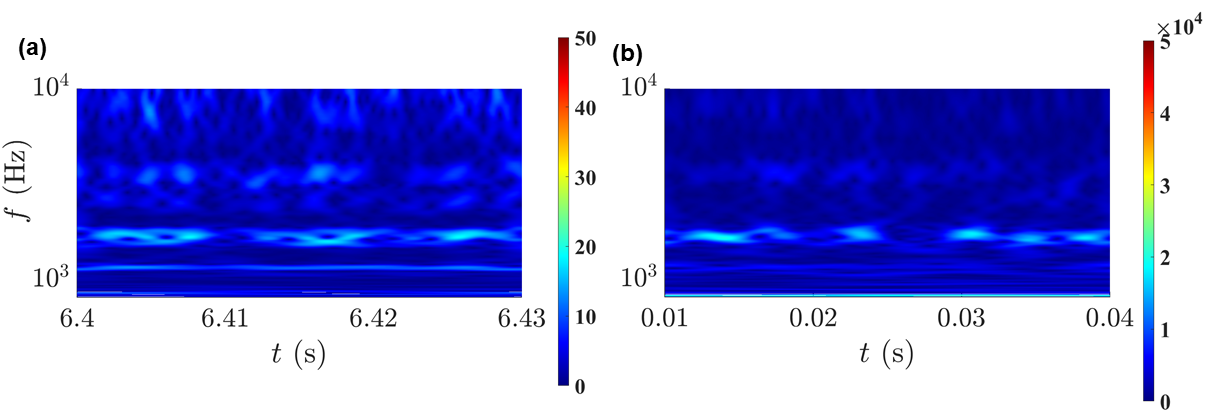}
\caption{(a) and (b) show the scalograms (frequency vs. time) of experimentally and numerically (CFD) obtained $p^{\prime}$ at 180.34 mm (c1 $\&$ c2) downstream from the dump plane or head end of the combustor. A dominant 1L mode, which is clearly observed during thermoacoustic oscillations near the dump plane region, almost diminishes in the downstream direction of the combustor.}
\label{fig_scalogram_downstream}
\end{figure*}

We perform the power spectrum of the acoustic pressure time series obtained at 180.34 mm downstream of the dump plane near the wall region (location of pt6 pressure transducer). We find that the 1L mode almost diminishes there. However, the bulk modes are present in that region (Fig.~\ref{fig_scalogram_downstream} (b)). 

\subsection*{B. Mutual influence between acoustic pressure and heat release rate oscillations towards a local minima of acoustic pressure fluctuation}\label{Appendix:B}

\begin{figure*}[ht!]%
\centering
\includegraphics[width=0.78\textwidth]{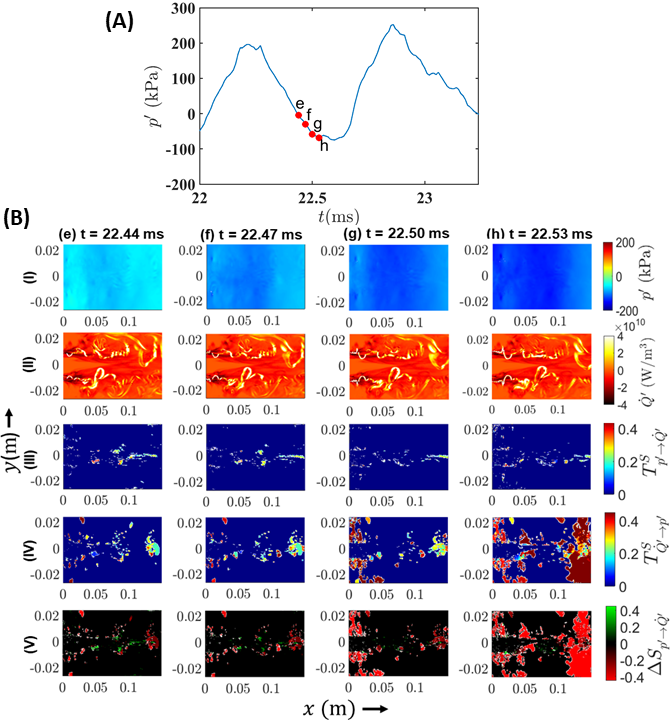}
\caption{A. Temporal evolution of $p^{\prime}$ is shown during 22 to 24 ms. we perform the analysis at four time stamps marked by red circles. 
B. The spatiotemporal distributions of $p^{\prime}$ (row-I) and $\dot{Q}^{\prime}$ (row-II) at 0 $\le x\le$ 0.15 m, -0.25 m $\le y\le$ 0.25 m and $z=0$ are presented in the first two rows. During $t = 22.44 - 22.53$ ms, we find that $p^{\prime}$ goes below zero and thus, reaches at local minima, representing the acousting damping during thermoacoustic instability. In this analysis, we interest to calculate the mutual influence between acoustic pressure and heat release rate fields when $p^{\prime}$ approach to the minima (negative peak). We can identify a few acoustic influencing zones along the center of the combustion chamber (see row-III, $T^{S}_{{p^{\prime}}{\rightarrow}{Q^{\prime}}}$). In contrast, the combustion influencing zones can be identified to be more spread over the dump regime at t = 22.50 ms (see row-IV, $T^{S}_{{Q^{\prime}}{\rightarrow}{p^{\prime}}}$). Further, the directionality index ($\Delta S_{{p^{\prime}}{\rightarrow}{Q^{\prime}}}$) is used to identify a distinct boundary between the direction of dominant interaction between heat release rate and acoustic fields prior to acoustic damping at row-V of the figure. The dimension of the combustor is shown in meter (m).}
\label{fig_STE_simu_pq_acoustic damping}
\end{figure*}

\begin{figure*}[ht!]%
\centering
\includegraphics[width=0.7\textwidth]{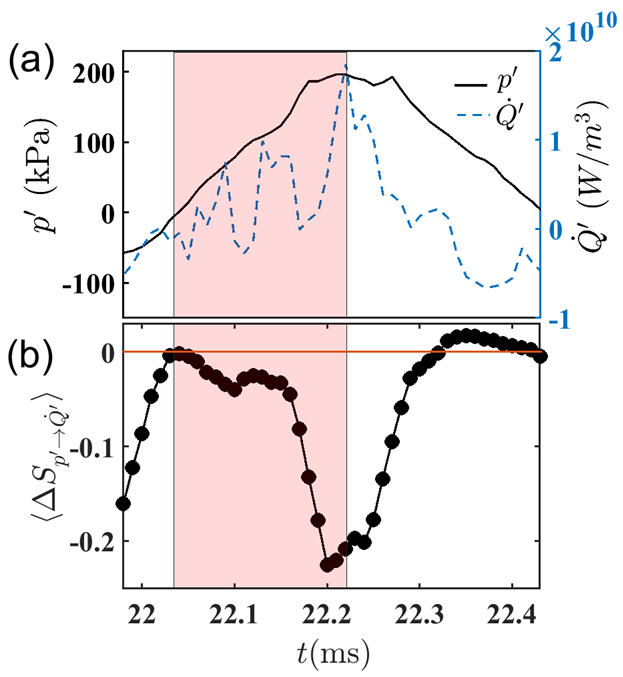}
\caption{We plot $p^{\prime}$ and $\dot{Q^{\prime}}$ in (a) during 21.9 to 22.43 ms. In (b), we estimate $\langle\Delta S_{p^{\prime}{\rightarrow}\dot{Q}^{\prime}}\rangle$ for the same time period. During this period, similar to the earlier observation, we find that $\langle\Delta S_{p^{\prime}{\rightarrow}\dot{Q}^{\prime}}\rangle$ becomes negative when $p^{\prime}$ rises at the dump region, confirming a stronger influence of $\dot{Q}^{\prime}$ on $p^{\prime}$ than $p^{\prime}$ on $\dot{Q}^{\prime}$.}
\label{fig:STE_pq_diff_cycle}
\end{figure*}

In Fig.~\ref{fig_STE_simu_pq_acoustic damping} (B), we try to investigate $T^{S}_{{p^{\prime}}{\rightarrow}{\dot{Q}^{\prime}}}$ (row-III), $T^{S}_{{\dot{Q}^{\prime}}{\rightarrow}{p^{\prime}}}$ (row-IV) and $\Delta S_{{p^{\prime}}{\rightarrow}{\dot{Q}^{\prime}}}$ (row-V) during the time period (time stamps are marked in Fig.~\ref{fig_STE_simu_pq_acoustic damping} (A)) when $p^{\prime}$ becomes locally minimum. Here, we consider a small regime of the combustion chamber near the dump plane i.e., $0\le x\le$ 0.15 m, -0.25 m $\le y\le$ 0.25 m and $z=0$. To correlate our understanding from STE approach with the variation in acoustic pressure and heat release rate fields during acoustic damping of a TAI cycle, we  represent the spatiotemporal behavior of those fields in the first and second rows of Fig.~\ref{fig_STE_simu_pq_acoustic damping} (B). From the distributions of heat release rate oscillations, we notice that the intensity of the flame becomes low compared to that seen during the increase in $p^{\prime}$ towards a local maxima. Also, the shape of the flame close to inlet of the chamber changes at $t=22.50$ ms as the flow field changes (not shown here). However, from the spatial distributions of $T^{S}_{{p^{\prime}}{\rightarrow}{\dot{Q}^{\prime}}}$ during $t= 22.44$ to $22.53$ ms, we find that $p^{\prime}$ influences $\dot{Q}^{\prime}$ along the center of the combustor. The spatial distribution suggests that acoustic pressure to heat release rate interaction  effectively happens from the center of combustor to flame, although $p^{\prime}$  interacts with ${\dot{Q}^{\prime}}$ at many points in the combustion chamber at that time. In addition, we observe a very few ${p^{\prime}}{\rightarrow}{\dot{Q}^{\prime}}$ influencing 
%cells 
zones at the dump region at $t=22.53$ ms.          

On the other hand, the influence of ${\dot{Q}^{\prime}}$ on ${p^{\prime}}$ gradually spreads over the dump regime and also extends towards the downstream direction during the period (see $T^{S}_{{\dot{Q}^{\prime}}{\rightarrow}{p^{\prime}}}$ in Fig.~\ref{fig_STE_simu_pq_acoustic damping} (B) IV). Therefore, although the flame intensity reduces during the period, we find an emergence of influence in the direction of ${\dot{Q}^{\prime}}$ to ${p^{\prime}}$ in the combustor. Further, from Fig.~\ref{fig_STE_simu_pq_acoustic damping} (B) V, we can demarcate the regions where the predominate direction of feedback interactions is found either along ${p^{\prime}}{\rightarrow}{\dot{Q}^{\prime}}$ or ${\dot{Q}^{\prime}}{\rightarrow}{p^{\prime}}$ using $\Delta S_{{p^{\prime}}{\rightarrow}{\dot{Q}^{\prime}}}$. Thus, similar to the positive ${p^{\prime}}$ of TAI cycle (Sec.~\ref{subsubsec:acoustic-heat}), ${\dot{Q}^{\prime}}$ has an important role in influencing the acoustic damping (${p^{\prime}} < 0$) during TAI since $\Delta S_{{p^{\prime}}{\rightarrow}{\dot{Q}^{\prime}}}$ exhibits negative values across a large area (see red colored regions of $\Delta S_{{p^{\prime}}{\rightarrow}{\dot{Q}^{\prime}}}$ during $t=22.44$ ms to $22.53$ ms in Fig.~\ref{fig_STE_simu_pq_acoustic damping} (B)). However, it is noteworthy that ${\dot{Q}^{\prime}}{\rightarrow}{p^{\prime}}$ influencing regimes are more spread and strong over the dump plane during positive ${p^{\prime}}$ than acoustic damping.

\subsection*{C. Quantification of mutual influence between acoustic pressure and heat relase rate oscillations during 21.9 to 22.43 ms }\label{Appendix:C}

We consider a cycle of TAI during 21.9 to 22.43 ms to check the variation of $\langle\Delta S_{p^{\prime}{\rightarrow}\dot{Q}^{\prime}}\rangle$ when $p^{\prime}$ approaches a local maxima (Fig.~\ref{fig:STE_pq_diff_cycle}). Similar to Fig.~\ref{fig_STE_quantification}, we find that the directionality measure drops to a large negative value during this period (see the shaded region in Fig.~\ref{fig:STE_pq_diff_cycle} (b)).

\bibliography{manuscript.bib}% Produces the bibliography via BibTeX.

\end{document}